# United in Diversity? Contextual Biases in LLM-Based Predictions of the 2024 European Parliament Elections

Leah von der Heyde [1]*, Anna-Carolina Haensch [2, 3, 4], Alexander Wenz [5], Bolei Ma [2, 3]

[1] GESIS – Leibniz Institute for the Social Sciences, Germany
[2] LMU Munich, Germany
[3] Munich Center for Machine Learning, Germany
[4] University of Maryland, United States
[5] University of Mannheim, Mannheim Centre for European Social Research, Germany

*Corresponding author: firstname[dot]lastname[at]gesis[dot]org

## Abstract

"Silicon samples" based on large language models (LLMs) have been argued to serve as efficient alternatives to surveys of humans, assuming that their training data includes information on human attitudes and behavior. However, LLM-generated survey data might exhibit biases, for example due to a lack of diversity in training and alignment processes, which risk reinforcing existing biases in research and policymaking. In this study, we examine to what extent LLM-based predictions of individual public opinion exhibit context-dependent biases by predicting the results of the 2024 EU elections. Before the elections have taken place, we prompt three LLMs with individual-level background information of 26,000 eligible European voters to predict each person's voting behavior. We show that when compared to the actual results, LLM-based predictions of *future* voting behavior largely fail, their accuracy is unequally distributed across *national* and *linguistic* contexts, and they require *detailed attitudinal information* in the prompt. The findings emphasize the limited applicability of silicon samples to public opinion prediction. In investigating their contextual biases, this study contributes to the understanding and mitigation of inequalities in the development of LLMs and their applications in computational social science.

## 1. Introduction

Large language models (LLMs) have recently emerged as a new tool for researchers in computational social science and have been proposed to complement existing methods for understanding human attitudes and behaviors. For example, research has started to assess to what extent LLM-generated synthetic data can be used as a viable and efficient alternative for collecting data about public opinion (e.g., Argyle et al. 2023; Bisbee et al. 2024; Dominguez-Olmedo et al. 2024; Sanders et al. 2023). Since LLMs are trained on large amounts of human-generated text data, their output has been argued to reflect human attitudes and behaviors. Thus, researchers hope that by conditioning an LLM with specific individual-level information in the input (so-called "persona prompting"), the LLM could be prompted to respond from that individual's perspective and – if scaled up – synthesize public opinion data for entire human samples. Such "silicon samples" have been proposed to ease the collection of previously unobserved public opinion data, including, but not limited to, data about hard-to-survey populations, sensitive topics, or future outcomes, and to facilitate questionnaire pre-testing and pilot studies. For picking up rapid shifts in public opinion, they promise to be quick, cheap data sources. As a result, an increasing number of enterprises offer survey and market research "solutions" based on silicon samples (e.g., Delve AI n.d.; Synthetic Users n.d.; Aaru n.d.; Levanti and Verret 2024).

However, previous research has found biases in LLM output with regard to various outcomes, including political attitudes and psychological measures (Sanders et al. 2023; Atari et al. 2023; Durmus et al. 2024; Santurkar et al. 2023; Von der Heyde et al. 2026; Kim and Lee 2023; Lee et al. 2023; W. Wang et al. 2024). Among the potential reasons for why these biases occur, unrepresentative training data regarding linguistic, social, political, and digital contexts are often mentioned (e.g., Von der Heyde et al. 2026). With more than 50% of website content language estimated to be English, the amount of available native-language training data for LLMs is considerably smaller for countries with any other native language. Moreover, the relationship between societal and political structure and public opinion formation differs between countries, and is likely not sufficiently represented in LLM training data. Finally, the training data is likely affected by coverage bias caused by the "digital divide". There may be differences between the respective target population and those who contributed to the specific texts selected for training LLMs. While it is difficult to identify their exact causes due to LLMs being "black boxes", such biases can challenge the validity of findings based on silicon samples, and risk reinforcing existing biases in social science research, policymaking, and society. Therefore, computational social scientists need to investigate if and under which conditions silicon samples can be applied for public opinion prediction by comparing different linguistic, political, social, and digital contexts.

Initial studies that used silicon samples for estimating public opinion, particularly vote choice, yielded results matching survey data in the context of the U.S. general population (Argyle et al. 2023; Kim and Lee 2023; Lee et al. 2023). Other research, however, has challenged these initial findings, particularly in other national contexts (Durmus et al. 2024; Qu and Wang 2024; Motoki et al. 2023) and languages (Von der Heyde et al. 2026; Qi et al. 2025). These studies find evidence for politically left-leaning, and culturally and linguistically WEIRD (Western, Educated, Industrialized, Rich, Democratic) biases in LLM outputs, reproducing simplified stereotypes for other national and linguistic contexts. However, existing cross-national studies on attitudinal biases in LLMs have typically employed *country-level* prompting only (Atari et al. 2023; Durmus et al. 2024; Motoki et al. 2023; Hartmann et al. 2023), not individual-level personas sourced from survey data. Such generic, country-level input only allows for generic output, not testing LLMs' capabilities in producing estimates of public opinion based on nuanced, individual-level predictions which might potentially result in better aggregate results. Previous research that used silicon samples based on *individual-level* characteristics for estimating public opinion, in turn, has identified biases regarding certain subgroups, but, thus far, has mostly been conducted in isolated national settings, not making any cross-national comparisons (Bisbee et al. 2024; Sanders et al. 2023; Santurkar et al. 2023; Von der Heyde et al. 2026; Kim and Lee 2023; Lee et al. 2023). An exception is the study by Bisbee et al. (2024), which finds in a supplementary cross-national analysis that ChatGPT performs similarly poorly across countries, with a tendency to predict attitudes that are more common in the benchmark survey data. The study does not test whether performance is related to the prompt language (they only used English) or content. Furthermore, Qi et al. (2025) compared persona-based estimations of U.S., German, and Chinese voting behavior simulated by GPT-3.5-Turbo to election studies in the respective countries. The results suggest that performance is better in English-speaking countries and two-party systems. Similarly, Qu and Wang (2024) assessed ChatGPT's performance in simulating the political attitudes of six global populations, finding that it more closely matched survey data for the three English-speaking populations, particularly the U.S., and that it exhibited biases towards demographic subgroups. While these case studies support the argument of a U.S. and English bias, they largely neglect an in-depth analysis of LLM performance differences between the other contexts.

Additionally, while most existing studies employed silicon samples for "predicting the past" (Argyle et al. 2023; Bisbee et al. 2024; Durmus et al. 2024; Qi et al. 2025; Qu and Wang 2024; Zhang et al. 2024; but see Kim and Lee 2023; S. Jiang et al. 2024), it is essential to assess their performance in making predictions of unobserved outcomes, such as future election results.[1] Such an investigation can illustrate how past training and survey data informs the prediction of future outcomes and how much past information is necessary for accurate prediction. Indeed, if detailed and timely individual-level survey data is necessary to make somewhat accurate predictions with LLM-generated samples, such samples may not be of much use to researchers, as they would still need to resort to surveys and possibly could even ask about the unobserved outcome of interest directly in those surveys.

In this paper, we aim to bridge these gaps between cross-national, individual-level, and future-outcome investigations and applications of silicon samples. Using the example of LLM-based predictions of voting behavior in the 2024 European Parliament elections, we examine to what extent LLM-based predictions of individual public opinion exhibit context-dependent biases by addressing the following research questions:

**RQ1.** Can LLMs predict the aggregate results of future elections?
**RQ2.** How does LLMs' predictive performance differ across countries?
**RQ3.** How does LLMs' predictive performance differ across prompt languages?

[1] Other LLM-powered election *pre*dictions do not rely on persona prompting, but on direct simulations of distributions (Bradshaw et al. 2024) or on LLM-based classifications of social media content (Cerina and Duch 2025; Gujral et al. 2024).

**RQ4.** How does LLMs' predictive performance differ depending on the information provided in the prompt?

Elections are a real-world example of an important, yet challenging prediction task in public opinion research. More specifically, the 2024 European Parliament elections provide a relevant test case for biases in LLM-based predictions across contexts, featuring both a comparable temporal and electoral reference point and high diversity in linguistic, social, political, and digital contexts across the 27 European Union (EU) member states. For an additional in-depth investigation of differences in LLMs' bias across languages, we select five countries, differing in native language Internet coverage, linguistic prevalence, language family, as well as in population size, geographic and political position within, and attitudinal position towards Europe. Following previous research (Argyle et al. 2023; Bisbee et al. 2024; Dominguez-Olmedo et al. 2024), we employ the silicon sampling approach: We sequentially prompt the LLM GPT-4-Turbo (OpenAI et al. 2023) with pseudonymized individual-level background information from an existing survey sample of approximately 26,000 eligible voters. For each individual, we create a description including socio-demographic and attitudinal information to prompt the LLM. Additionally, the profiles include information about the parties competing in the respective country. Prior to the European Parliament elections, we then ask the LLM to predict each person's voting behavior. Asro robustness checks, we perform the same analyses on two open-weight LLMs, Llama 3.1 (Dubey et al. 2024) and Mistral (A. Q. Jiang et al. 2023), using the same prompts and model configuration for input, and compare the LLM-generated estimates to estimates based on post-election surveys.

Applying the silicon sampling approach to election prediction, we show (1) how well popular LLMs perform at predicting *future* voting behavior based on past training and *individual-level* survey data, (2) how this performance differs *across national* and *linguistic* contexts, and (3) whether it is currently feasible to supplement survey data with LLM-based data given limited individual-level *information* provided in the prompt. In investigating the contextual differences of LLM-based predictions of public opinion, our research contributes to the understanding and mitigation of biases in LLM applications in computational social science.

## 2. Data and Methods

### 2.1 Sample and LLM selection

To examine biases of silicon samples in a variety of linguistic, social, political, and digital contexts, our study spans all 27 EU member states (EU-27). For an additional in-depth investigation of differences in LLMs' predictive performance across languages, we select five countries differing in native language Internet coverage (W3Techs 2024), linguistic prevalence within the EU, language family, as well as in population size, geographic and political position within, and attitudinal position towards Europe (for details, see Appendix III): France, Germany, Poland, Slovakia, and Sweden. Ireland serves as an English-language baseline.

To create a realistic sample of individual-level profiles on which we base our predictions of vote choice in the 2024 European Parliament elections, we rely on the most recent Eurobarometer survey data available before the elections, EB 99.4 from May-June 2023. This data has been collected with face-to-face interviews of EU citizens aged 15 years and over and resident in the EU-27, based on stratified, multi-stage probability samples (B. European Commission 2024). From this data, only voting-eligible EU citizens are selected, resulting in a sample of about n=1000 per EU member state (with the exception of Cyprus, Luxembourg and Malta, with a sample size of about n=500 each, and Germany, with about n=1500) or about 26,000 respondents in total. For summary statistics of all variables, see Appendix II.

Simulating a realistic use-case, we use one of the most popular and powerful LLMs at the time of the elections, GPT-4-Turbo (version 2024-04-09), which has a cutoff date in December 2023 (OpenAI n.d.). Compared to older GPT models, it is supposed to have better multilingual capacities, be better at solving complex instructions, and less likely to "hallucinate", that is, provide fabricated output. Finally, its performance in predicting public opinion was shown to be better when adding information beyond demographics (Lee et al., 2023), and in different languages (P. Wang et al. 2024). To understand whether any biases we find can be generalized across LLMs, and to guide the development of future LLMs, we perform the same analyses on two open-weight LLMs, Llama 3.1 (knowledge cutoff December 2023, Dubey et al. 2024) and Mistral (knowledge cutoff December 2022,

A. Q. Jiang et al. 2023). Llama is optimized for multilingual dialogue use cases and supposed to be comparable to GPT but superior to other open source LLMs. Mistral is chosen for its robust performance in handling a wide range of tasks, including those requiring content reasoning and creative writing, which was shown to complement and in some cases even surpass the strengths of Llama models (MistralAI 2023).

### 2.2 Prompt creation

For each individual in the Eurobarometer sample, we create a description including socio-demographic and attitudinal information with which we prompt the LLMs using second-person pronouns (Bisbee et al. 2024). Prompts vary with regards to citizens' age, gender, education, socio-economic class, occupation status, and urbanity. Additionally, the profiles include information about individuals' political interest, ideology, trust in the EU, and attitude towards European integration. These variables have been identified as determinants of voting behavior in EU elections (Braun and Schäfer 2022; Ford and Jennings 2020; Giebler and Wagner 2015). In using vote choice as a test case for examining biases in predictions based on silicon samples, we simulate a realistic use case where researchers and practitioners with limited resources and *limited information* about their target population rely on *off-the-shelf LLMs*. The aim of the study is to assess the quality and *systematic differences* of LLMs' predictions across contexts when *holding information constant*. Thus, we do not account for country-specific determinants of voting behavior in European elections to ensure cross-country comparability of the prompts, even though this might limit the LLMs' predictive accuracy. Finally, the prompts feature the parties competing in the respective country that a) currently have a seat in the European Parliament or b) polled above the respective country's electoral threshold at the time of data collection (for details on which parties these are, see Appendix I). For countries that do not have an electoral threshold in EU elections, we require a minimum of 2%, as this is the minimum threshold all countries have to implement by the 2029 EU elections (Sabbati and Grosek 2023). We also include the pan-European party Volt for countries in which it is competing. The existence of parties below this threshold is indicated by "several smaller parties" at the end of the list. The order of parties is randomized to avoid any tendencies by GPT models to choose (one of) the first option(s) listed (Brand et al. 2024; Pezeshkpour and Hruschka 2024). However, the order of variables, that is, sentences in the prompt, is not randomized, as this would risk contrived language that could impact the predictive accuracy and because assessing the impact of prompt engineering on prediction quality is not the aim of this study. Missing values on any of the individual variables are imputed for 6800 respondents (26 % of the total sample) using multivariate imputation by chained equations (Van Buuren 2018). Clarifying the year and that the individual is voting-eligible aims to avoid erroneous predictions due to assumed ineligibility (Von der Heyde et al. 2026). In order to avoid refusals by the LLMs and therefore incomplete output, we include a request to make the best possible prediction based on the data provided. Finally, the instruction to keep the answer short aims to ensure cost-efficiency, as the LLMs might otherwise unnecessarily elaborate on their choices. Figure 1 shows an exemplary prompt. For details on the prompting variables and their Eurobarometer source, see Appendix I.

For the entire sample spanning all EU member states, we prompt the LLMs with the individual profiles in English, once containing only socio-demographic information, and once also containing the attitudinal variables (underlined in Figure 1). Additionally, for profiles from the five selected countries, we prompt the LLMs in the respective country's native language. For translations of the prompts, performed by researchers native in the respective language and on the basis of the native version of the questionnaire, see Appendix I.

> *The year is 2024. You are a voting-eligible citizen of an EU-member state living in* ***Germany****. The parties competing in the 2024 elections to the European Parliament are* ***CDU/CSU, SPD, Bündnis 90/Die Grünen, FDP, Die Linke, AfD, Volt, Freie Wähler, Tierschutzpartei, Familienpartei, Bündnis Deutschland, Bündnis Sahra Wagenknecht, Piratenpartei****, and several smaller parties.*
> *You are* ***29*** *years old. You are* ***female****. You have a* ***university*** *education. Economically, you are* ***upper middle*** *class. You are* ***working****. You live in a* ***big city****. You are* ***very*** *interested in politics. Ideologically, you are* ***center-left****. You* ***think*** *that more decisions should be taken at the EU level. You tend* ***to trust*** *the European Union.*
> *Will you vote in the 2024 elections to the European Parliament, and if so, for which party? Use the information above to make as good a prediction as possible, and keep your answer as short as possible, if possible just "No" or the name of the party you would predict.*

*Figure 1: Example prompt (prompt variables in bold; attitudinal information underlined).*

## 2.3 LLM configuration

We automate the data collection through the Azure OpenAI REST API for GPT (OpenAI et al., 2023), and through local instances of Llama and Mistral (Meta n.d.; MistralAI n.d.). Azure OpenAI provides private cloud inference with contractual data-processing guarantees, thereby ensuring the input data is not passed on to third parties (i.e., OpenAI servers). Open-weight LLMs, in contrast, typically can be downloaded and run on local computing infrastructure to begin with, minimizing privacy concerns. We employ zero-shot prompting and, in line with previous studies (Aher et al. 2023; Bisbee et al. 2024; Lee et al. 2023; Tjuatja et al. 2024), configure the LLMs to a temperature of 0.9. To further control the LLMs' completions' length, we limit the output to a maximum of 40 tokens, which allows for a response including a complete sentence with all necessary information in all target languages. As previous research showed little variance in individual vote choice predictions when prompting GPT repeatedly (Von der Heyde et al. 2026), we only prompt the LLMs once per individual. We generate the data shortly before the European elections are held (June 6-9, 2024), between May 29 and June 4, 2024. Data for the robustness checks was generated on July 29 (Llama) and August 1 (Mistral) – however, as the LLMs' knowledge cutoffs are before the elections took place, this should not impact the results.

## 2.4 Vote choice extraction

Vote choices are extracted from LLM completions based on a set of predefined keywords per competing party, as well as non-voting and invalid voting (see Appendix I). As European elections typically feature a large number of very small political parties beyond the ones established in national politics, votes for parties that do not meet the respective country's electoral threshold in the official results (Sabbati and Grosek 2023), or, in cases of no threshold, parties that do not obtain a seat in the newly elected parliament (European Parliament n.d.), are summarized as "Other" for the analyses that follow. As this study aims to depict a realistic use-case as opposed to optimizing predictions a priori, completions that do not contain a definite party choice are recorded as missing and only counted for turnout calculation if the prediction clearly states the person would have voted, but not for vote share calculations (see Appendix II for proportions of missing values).

### 2.5 Analysis

We weight the extracted results with the Eurobarometer-provided weights to better approximate the target population. To answer our first research question, we compare the aggregate predicted voting behavior when prompted in English to the official national-level results across all 27 countries, differentiating between turnout and party vote shares among voters. Specifically, we compare the mean and variance of predicted and actual turnout, as well as several metrics for correct party vote share prediction, including prediction of the winning party, the rank ordering of parties, and average absolute differences in party vote shares per country as well as across European parliamentary groups as announced in the post-election constitutive session (European Parliament n.d.). We do not account for the different electoral systems in place in the different countries, nor for the different electoral thresholds, both of which impact voting behavior and vote aggregation, thereby potentially limiting the predictive accuracy. Future research could investigate whether such adjustments improve predictions.

To tackle our second research question, we contrast the differences in predicted turnout and party vote shares within the EU-27. We compare countries according to whether they have compulsory voting, their European region (EuroVoc n.d.), and their language family (Wikipedia, 2024). In line with our third research question, we also analyze the LLMs' predictive performance based on prompts in English and the five selected countries' native languages. Regarding our fourth research question, we compare whether predictions containing the full set of information in the prompt perform better than those based solely on socio-demographic information.

As an additional sensitivity analysis, we compare our results to estimates obtained via the Eurobarometer post-election survey from June-July 2024 (EB 101.5, European Commission and European Parliament, Brussels 2026) using the same filtering (voting-eligible EU citizens) as with the LLM-based data and using the weights provided in the dataset. To assess which estimate is closer to the real results, we calculate the (mean) absolute errors of either source – LLM or post-election survey – to the real election results, differentiating between turnout and vote choice, by country and region. From this, we calculate the difference in difference between the two data sources and the real results. We also calculate the magnitude of differences between the LLM and post-election survey data estimates.

In our analyses, we do not report confidence intervals or conduct traditional tests for statistical inference. Doing so would imply that the primary source of error stems solely from the sampling of observations – as is typically assumed in traditional survey research and related studies – even when LLM outputs are prompted using information derived from a sample. Instead, we argue that additional sources of error can arise from biases inherent in LLMs' data-generating process.

Data collection and analysis is conducted using the software R, version 4.3.2 (R Core Team 2024), especially the packages *tidyverse* (Wickham et al. 2019), *mice* (Van Buuren and Groothuis-Oudshoorn 2011), *rgpt3* (Kleinberg 2024), and *survey* (Lumley 2004).

## 3. Results

### 3.1 Overall prediction of EU election results

Despite the capabilities of GPT-4-Turbo, we are still far from being able to use it as an accurate and reliable prediction tool for public opinion: Predictions of turnout and party vote shares in the 2024 European elections based on silicon samples of the voting population fail. With an average predicted turnout of 83%, predictions based on GPT-4-Turbo overestimate turnout by 34 percentage points on average, not capturing the substantial variation between countries (Figure A1). Predictions of turnout almost all range above the total range of actual turnout. Considering party vote shares, GPT-4-Turbo-based predictions mostly fail to predict the winner (11 out of 27 correct) or ranking of parties (Figure 2), with the LLM only identifying 8% of party ranks correctly on average (with a median of 0%). Predictions of individual party vote shares often differ greatly from the actual result (Figure 3), with average differences of five to nearly 17 percentage points per country. This average per country masks a high variation between parties, with larger differences between predicted and actual vote shares especially for parties not belonging to the Green or Left parliamentary groups (Figure A2), confirming findings from previous research and including the two biggest groups left and right of center and the newly formed right-wing groups. The former have suffered substantial national losses in recent years, which may not have been picked up by GPT-4-Turbo due to the temporal limits of its training data.

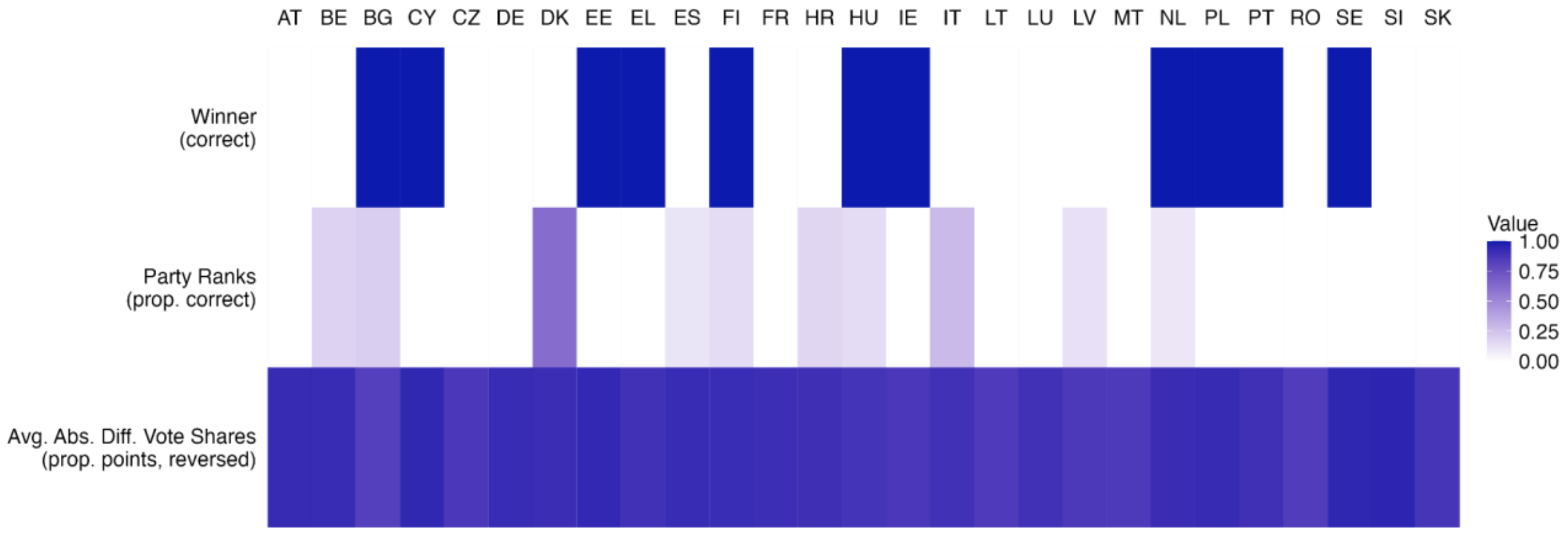

Figure 2: *Predictive performance of GPT-4-Turbo for the 2024 EU election party results (based on full English prompt).*

Note: *Average absolute differences in vote shares have been reversed so that higher values correspond to better predictive performance in line with the other metrics. Example: an average absolute difference of 5 percentage points (0.05) would be displayed as 0.95.*

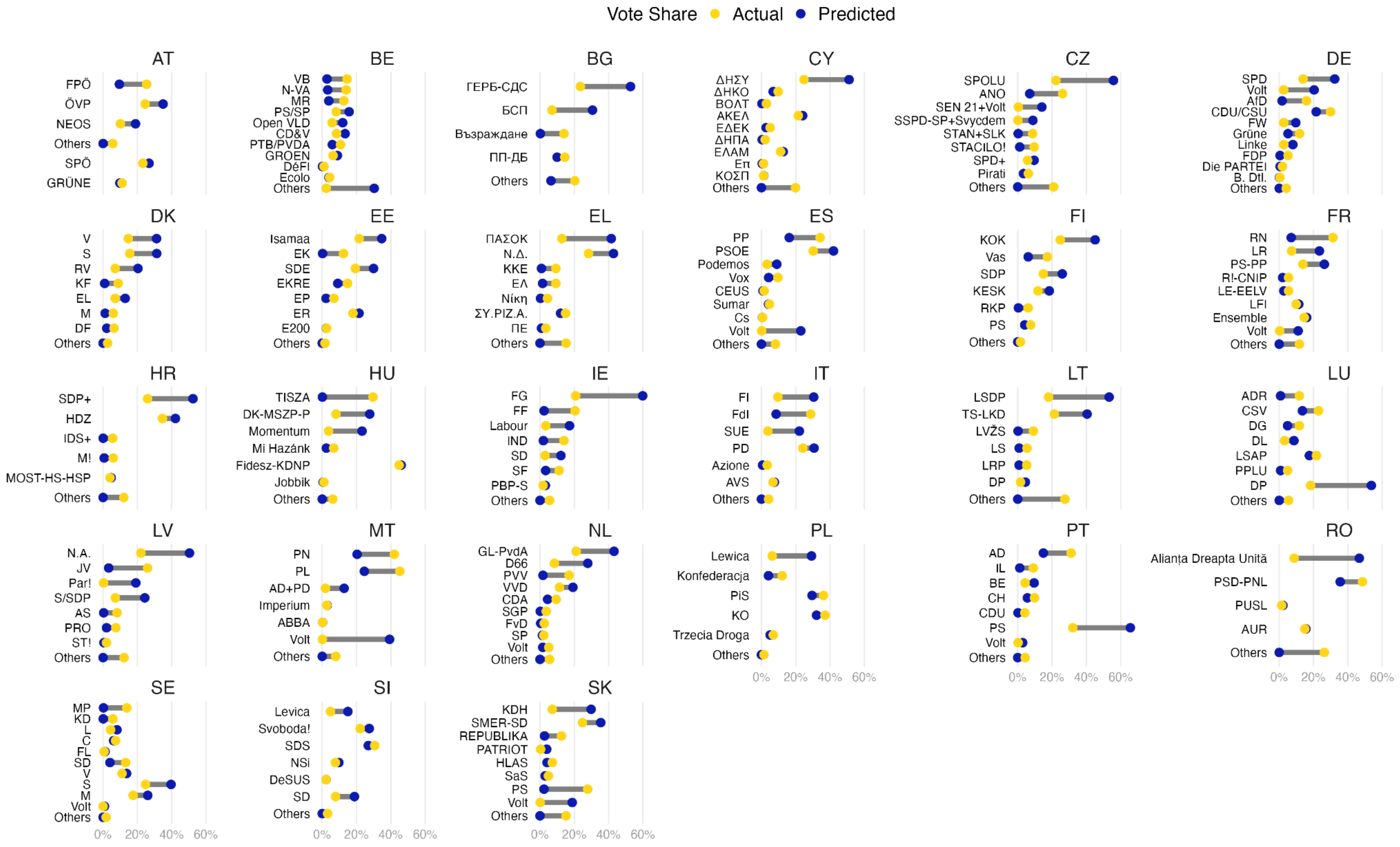

Figure 3: *Differences between actual party vote shares in the 2024 EU elections and and GPT-4-Turbo's predictions, by party (based on full English prompt).*

### 3.2 Differences in predictive performance across countries

For English prompting overall, GPT-4-Turbo's predictive performance of turnout is higher for countries with high actual turnout (Figure 4), while it overestimates turnout especially for countries with typically low actual turnout. This pattern can be explained by the LLM’s tendency to predict rather high turnout regardless of country, and holds even for the four countries with compulsory voting. The difference between predicted and actual turnout is among the lowest for Belgium and Luxembourg, Western European countries with French as an official language, one of the most dominant languages in Europe. In contrast, for Greece and Bulgaria, which are situated in South-East Europe and whose native languages use cyrillic alphabets and are less commonly used, the differences are among the highest.

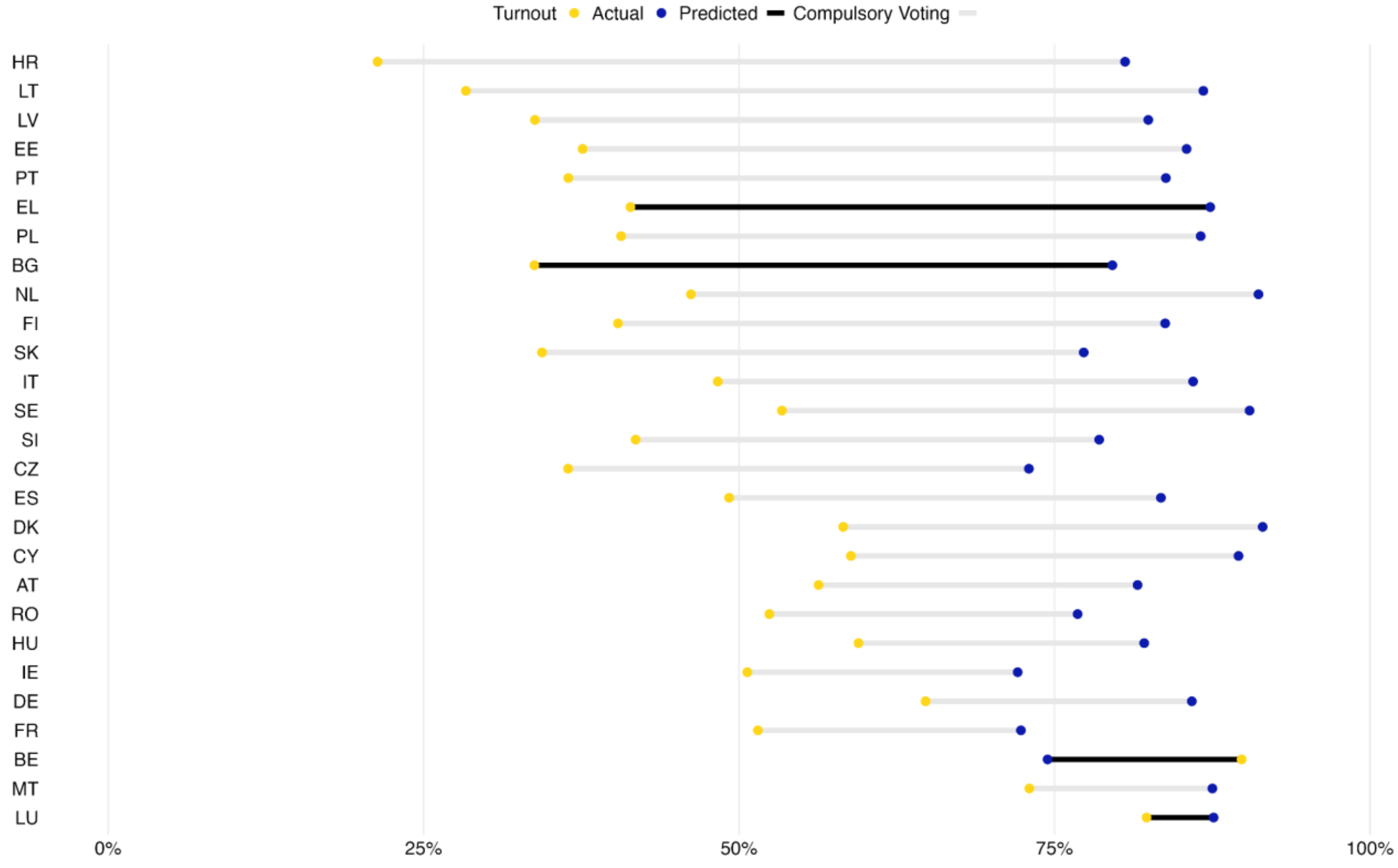

Figure 4: *Differences between actual turnout in the 2024 EU elections and GPT-4-Turbo’s predictions (based on full English prompt).*

When differentiating regions, the LLM’s overestimations of turnout tend to be higher for Eastern and Southern European countries, especially if considering the Baltic states (Lithuania, Latvia, and Estonia) as (historically) Eastern rather than (aspirationally) Northern European (Figure 5A). This pattern is confirmed when investigating native language families (Figure 5B): Overestimations of turnout are higher for Baltic and Slavic language countries when prompting GPT-4-Turbo in English. As especially Slavic languages are native to Eastern European countries, it is no surprise that these patterns overlap. Historically, turnout tends to be lower in non-Western European countries. However, the LLM is unable to capture this pattern, but assumes the high turnout levels of Western European countries that typically speak one of the more dominant Germanic or Romance languages, such as English, German, or French.

The same holds when it comes to predictions of party vote shares, which on average differ more from the actual results for Eastern and Southern European countries (again, especially when considering the Baltics as part of this group; Figure 5C/D) and such with Slavic or Baltic native languages. As evidenced by the case of Romania, a country with a Romance native language but among the countries with the highest difference between predicted and actual party vote shares, linguistic and geographical factors likely interact when it comes to GPT's predictive accuracy.

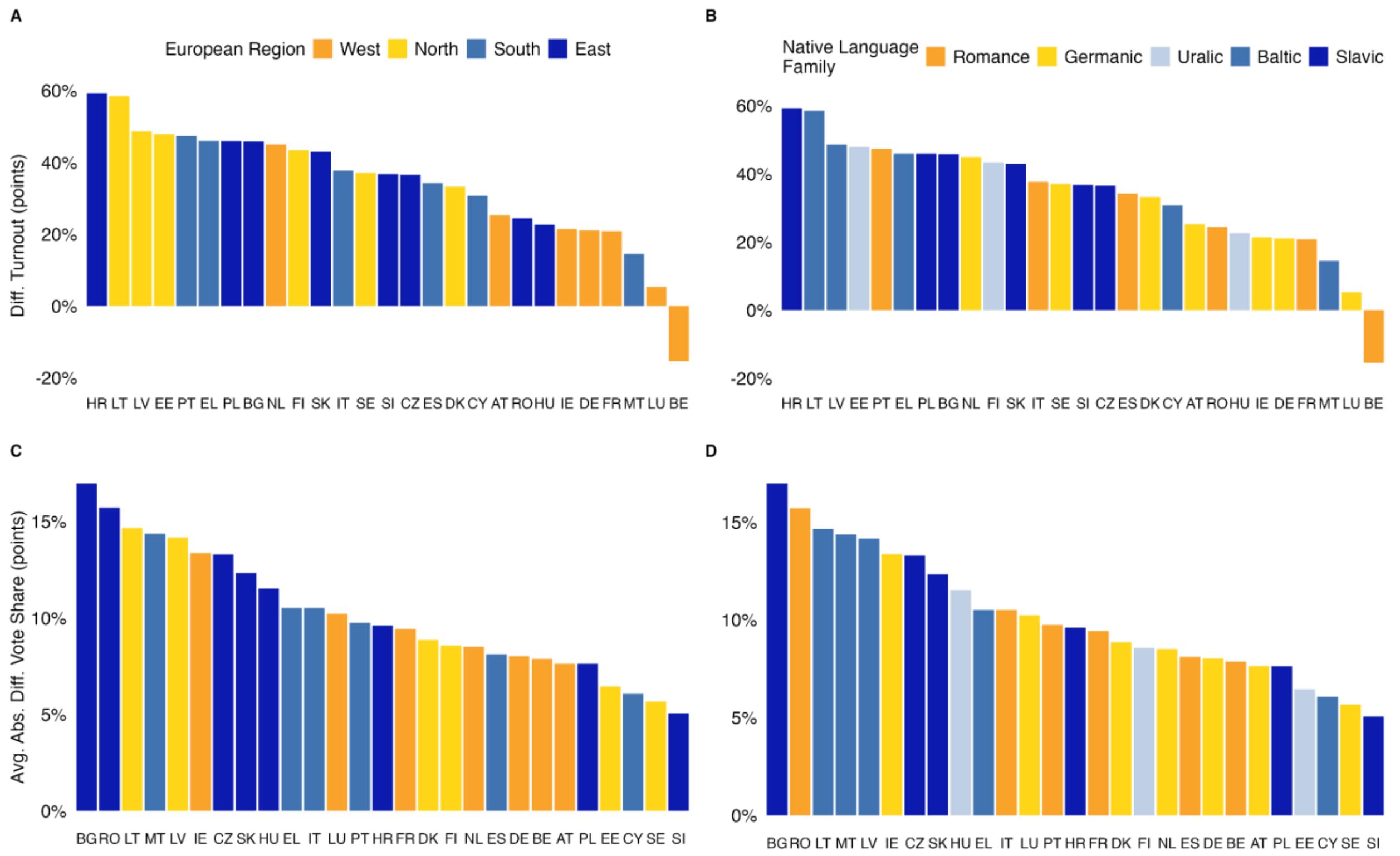

Figure 5: *(Average) difference between actual turnout and party vote shares in the 2024 EU elections and GPT-4-Turbo's predictions (based on full English prompt)* ***by region and language family****.*

### 3.3 Differences in predictive performance across prompt languages

In all of the five countries examined, prompting in the native language leads to an even bigger overestimation of turnout than when prompting in English (Figure 6A). The difference in difference of turnout estimation between English and native-language prompting is especially strong for France, followed by Slovakia. While there is barely a difference between prompt languages for Poland, the overestimation is particularly large regardless of language, at over 40 percentage points. When it comes to differences in party vote shares, the pattern somewhat reverses (Figure 6B). Notably, the average difference between predicted and actual vote shares is highest for the benchmark Ireland. This may be due to Ireland's complex single-transferable (ranked choice) voting system, which is not accounted for by the LLM or the aggregation.

To summarize, English-language prompting returns better predictions than native-language prompting for turnout (Figure 6A), but not as much for party vote shares (Figure 6B). The size of differences between English- and native-language prompting depends on the country in question, suggesting that GPT-4-Turbo is worse at predicting Eastern European voting behavior regardless of prompt language.

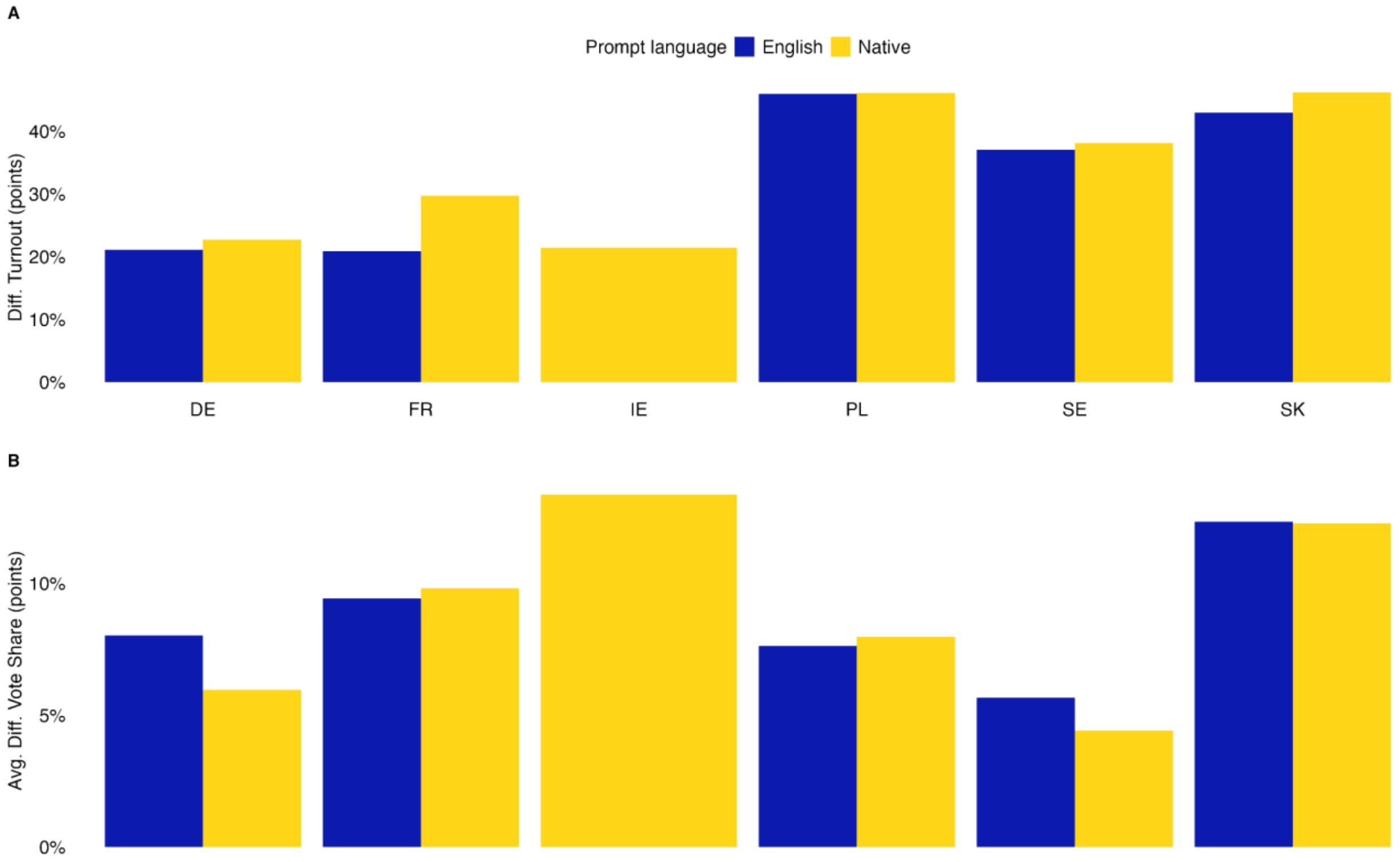

Figure 6: *(Average) difference between actual turnout and party vote shares in the 2024 EU elections and GPT-4-Turbo's predictions (based on full prompt)* ***by prompt language****.*

### 3.4 Differences in predictive performance depending on prompt content

When prompting GPT-4-Turbo in English with only demographic information about European citizens, the LLM tends to overestimate turnout even more (Figure 7A), and make even less accurate predictions of individual party vote shares than when prompted with additional attitudinal information for most countries (Figure 7B). Even in Belgium, where the full prompt led to an underestimation of turnout, GPT-4-Turbo overestimates turnout. For eight countries, predicted vote shares based on demographic information are closer to the actual result than those based on more detailed information. This includes Baltic states, some Eastern European countries, as well as Luxembourg and Malta. The apparent randomness of these results suggests an underlying randomness in when LLM-based predictions of voting behavior are correct, questioning the reliability of the method. Per-country-averages of absolute differences between predicted and actual vote shares for individual parties also have a lower variance when using only demographic information, suggesting that GPT-4-Turbo systematically misestimates vote shares regardless of the country or individual in question without additional information that would provide nuance (Figure A3). In other words, if provided with attitudinal information about individuals in addition to sociodemographics, GPT-4-Turbo's predictions of voting behavior are better.

Also when using native-language prompting, providing only demographic information leads to vastly higher overestimations of turnout compared to the full set of information (Figure 8A) and larger differences to actual party vote shares (Figure 8B). While the difference between demographic and full prompt is especially large for German and French, the level of divergence from the actual result is generally higher for Polish and Slovak (for turnout), suggesting a systematic bias against those countries and languages. GPT-4-Turbo is systematically worse at predicting voting behavior for Eastern European countries and/or countries with Slavic languages, regardless of prompt language or the amount of information provided in the prompt.

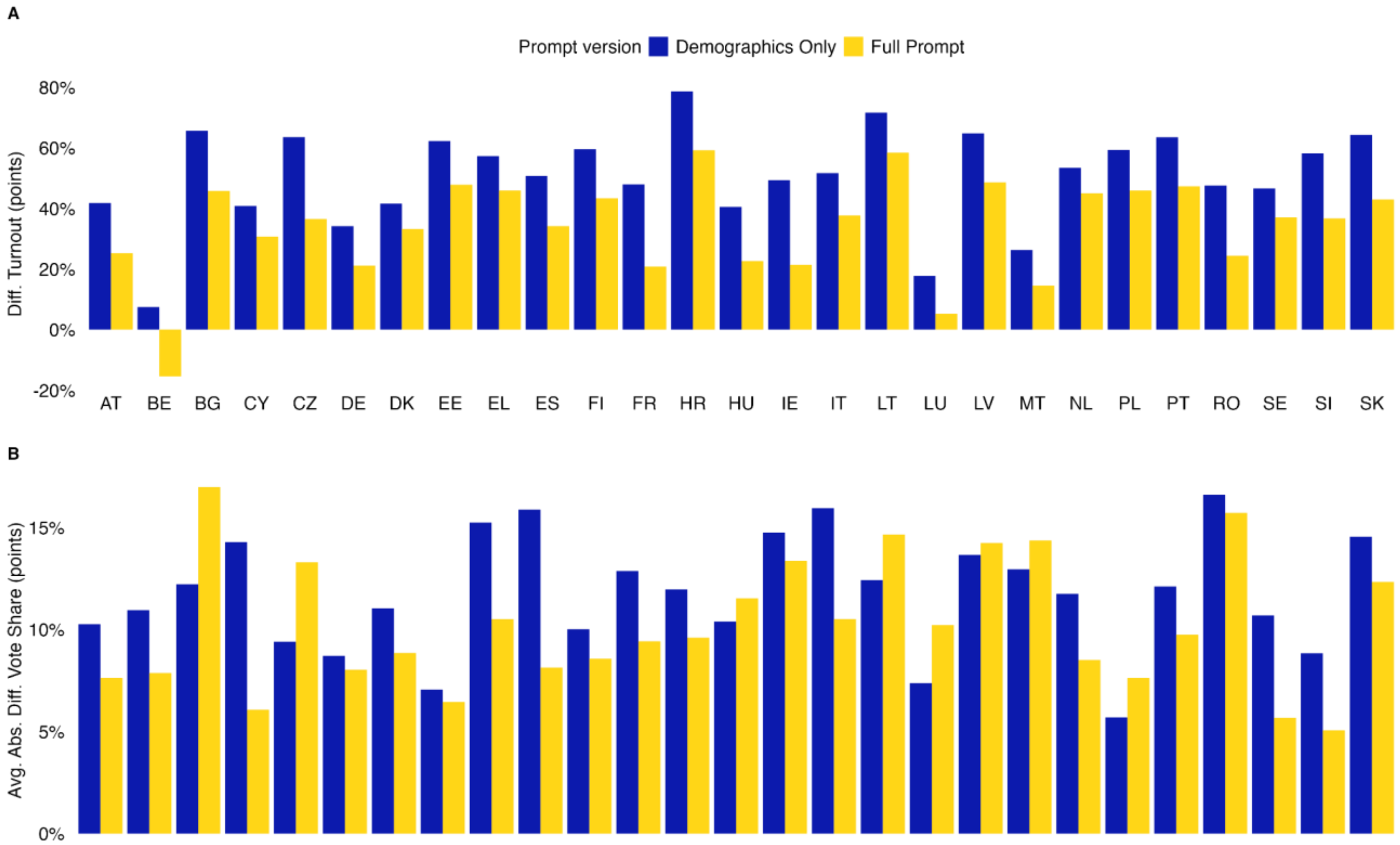

Figure 7: *(Average) difference between actual turnout and party vote shares and predictions using GPT-4-Turbo (based on* ***English*** *prompt)* ***by prompt content****.*

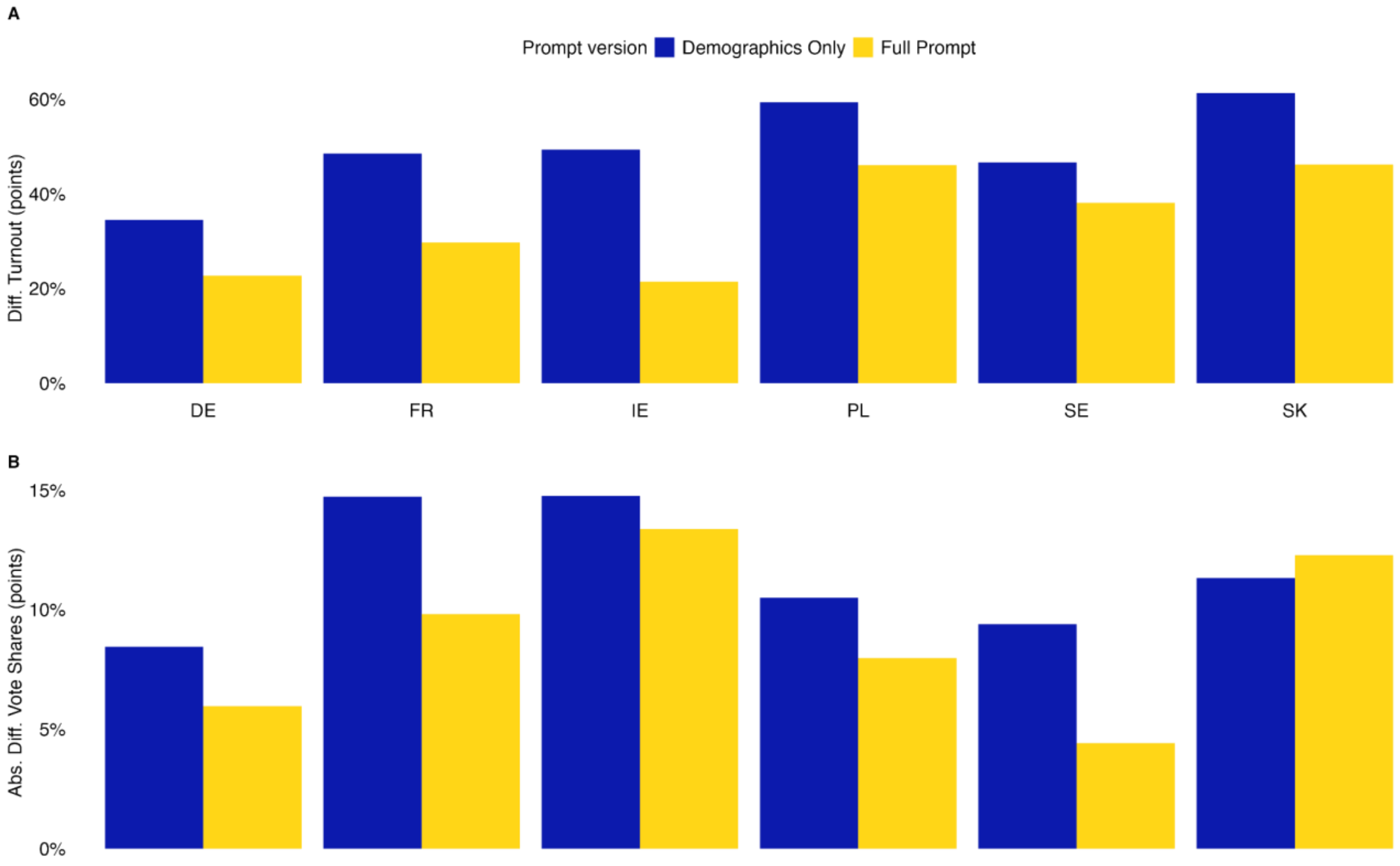

Figure 8: *(Average) difference between actual turnout and party vote shares and predictions using GPT-4-Turbo (based on* ***native language*** *prompt)* ***by prompt content****.*

### 3.5 Comparison to open-weight models

Using the same prompts and model configuration, the open-weight models Llama 3.1 and Mistral appear even less suitable for predicting public opinion based on silicon samples. Compared to GPT, predictions using Llama 3.1 lead to larger overestimations and bigger contextual biases when it comes to turnout (see Appendix III for figures). However, Llama-based predictions are not as biased when it comes to vote shares (although similarly wrong on average). The same holds for predictions based on only demographic information, which are much worse for turnout, but not so much for vote shares. Llama exhibits even poorer predictive performance when prompted in native languages, which can be attributed to its more limited multilingual capacities.

Llama-based predictions overestimate turnout in every country and to an even larger extent than GPT, predicting an average of 95% (an overestimation of 46 percentage points). Echoing GPT, predictions are better for countries with high actual turnout and those countries with compulsory voting that are Western European using a dominant language. Predictive performance of party popularity is similarly poor as GPT in terms of winning party, party ranking, and average absolute differences to actual party vote shares. Biases against Eastern European countries and countries with Slavic or Baltic languages are more pronounced for predictions of turnout than party vote shares when using Llama than when using GPT. Despite the difference in patterns when it comes to the outcomes investigated (turnout vs. vote shares), these results suggest that biases against Eastern European countries are present regardless of the brand of LLM used. In contrast to GPT, prompting in a country's native language yields mixed results when it comes to predictions of turnout, and predictions of vote shares based on English-language prompting outperform those with native-language prompting for all countries, especially France and Poland. Overall, this suggests that Llama's multilingual capacities are not as good as GPT's. While Llama has been trained on a broad range of languages, from our selection, only English, French, and German are officially supported use cases (Meta, n.d.). Thus, it is not surprising that its performance is weaker. Finally, providing only demographic information about individual voters leads to an even larger overestimation of turnout in all countries and higher divergences of predicted compared to actual party vote shares in most countries. While this pattern holds for native-language prompting regarding turnout predictions, the divergence to actual vote shares is not much different with or without attitudinal information.

Mistral largely did not follow the instruction of keeping the answer as short as possible, but instead either repeated the information contained in the prompt or stated that it was too difficult to make a definite prediction, both resulting in a disproportionate amount of completions lacking a vote choice, that is, missing values (see Appendix III for details). Analyses based on the remaining data would neither be meaningful nor comparable to the other models. We conclude that Mistral cannot be used for generating silicon samples for public opinion prediction in a similar manner as other models.

### 3.6 Comparison to post-election survey data

Our LLM predictions are based on the most recent publicly available pan-European probability survey data at the time of the elections and LLM data generation – a Eurobarometer survey from a year before the elections took place. Since then, more data has been released. We thus horserace the results of our “best”-performing configuration (GPT-4-Turbo with the full-information prompt in English) to estimates of voting behavior based on survey data collected just after the elections, from the same survey, with the same target population (voting-eligible EU citizens) and survey weights.

Discrepancies between LLM and survey estimates are large. On average, the LLM predicts a thirteen percentage point higher turnout than the post-election survey. The per-country *average* difference in vote shares ranges between four (Spain) and 32 (Romania) percentage points (details in Appendix III). The discrepancies are especially large for Eastern European and Baltic countries – exactly those countries where GPT failed the most.

Importantly, when comparing both data sources to the actual election results via difference-in-difference, we find that the post-election survey is closer to the results for a large majority of countries – regarding both turnout and party preferences.[2] Once again, the differences in performance are particularly large for Eastern European countries (Figure 10). Where GPT shows the largest deviations to the results, the survey data does not.

This confirms our main results: Not only do the LLM predictions diverge substantially from election *and survey* results, they do so systematically more for certain countries. Even when disregarding different electoral systems or electoral thresholds, post-election survey reports still outperform the LLM predictions based on pre-election samples.

[2] However, for most countries, both data sources overestimate turnout. For this and detailed per-party vote share differences, see Appendix III.

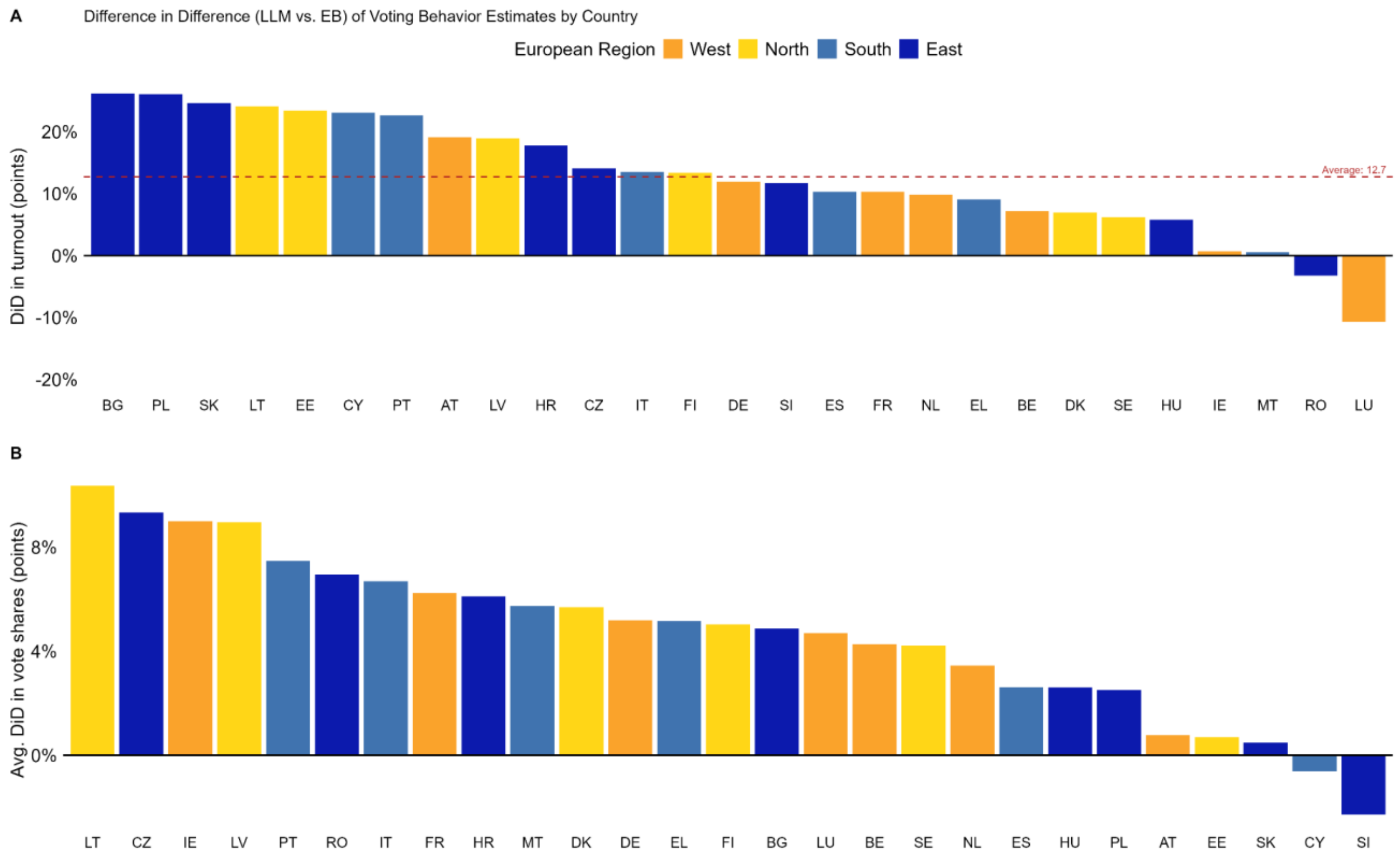

Figure 10: (Average) *difference in differences of GPT-4-Turbo (full, English prompt) and post-election Eurobarometer estimates to the real election results – turnout (panel A) and vote shares (panel B)* ***by region****, using absolute differences: DiD = |LLM - real| - |EB - real|. A positive "difference in difference" indicates that the post-election survey estimate is closer to the real result, a negative one indicates that the GPT prediction is closer.*

## 4. Discussion

Our results show that overall, LLMs fail at predicting turnout and party vote shares in the 2024 European elections based on silicon samples of the voting population – they overestimate turnout and are largely unable to accurately predict the winner, rank ordering, or individual party vote shares. Only providing socio-demographic information about individual voters further worsens the results, casting severe doubts on the feasibility of using silicon samples as a supplement, let alone substitution, of detailed survey data. Finally, the LLMs are especially bad at predicting voting behavior for Eastern European countries and countries with native Slavic languages, regardless of language used or the amount of information provided in the prompt, suggesting systematic contextual biases.

Predicting political attitudes and behaviors in multi-party contexts is more complex than in the U.S. two-party system (Von der Heyde et al. 2026), which most previous studies on silicon sampling investigated. As our findings show, predictions of *future* public opinion based on *off-the-shelf* LLMs do not live up to the hope of being a resource-efficient alternative in just any context, as they are not able to capture the complex mechanisms behind public opinion formation equally across contexts if these mechanisms are not featured in the training data (McCoy et al. 2024). Considering *what purpose* LLMs were trained to fulfil along with *how* they were trained to fulfil it (McCoy et al. 2024) and the training's *temporality* can help explain why LLMs fail in this task. Previous research has found that LLMs are better at retrodiction, that is, retroactively imputing past opinions, than at predicting attitudes on new survey items, policy issues or events that occurred past its training data. When predicting, LLMs seem to instead generalize along broad ideological lines without regard for nuance (Sanders et al. 2023; Kim and Lee 2023; Barrie and Cerina 2026; Gordon et al. 2026) – something that is reflected in the response distributions, which are different from human-generated survey data, often being less diverse (Bisbee et al. 2024; Dominguez-Olmedo et al. 2024; Hämäläinen et al. 2023). This should not be surprising, considering LLMs were trained to predict the most likely next words following a string of words (Resnik 2025). Put differently, they will output words that follow previous words with a high probability based on their training corpus. In instances where information about the task (here, predicting a specific, unobserved election outcome) occurs in the training data with low probability, LLMs will output words that, in its training corpus, are related to cues in the input prompt with a high probability, even though the context of the input might be a completely different one (McCoy et al. 2024). Furthermore, traditional polls, whether regarded as input knowledge sources or output benchmarks for LLMs, provide a snapshot in time, both in terms of a population's structure and its attitudes. Public opinion, however, is volatile, and while voting behavior has certain stable long-term predictors (Rattinger and Wiegand 2014), it is susceptible to shock events. Such short-term contextual changes and ensuing shifts in party popularity and strategic voting cannot be captured by LLMs with a knowledge cutoff far ahead of the event they are supposed to predict. For example, ahead of the 2024 European elections, several scandals within the far-right parliamentary groups dominated the news cycles and debates. In the specific case of elections, comparing the LLM-predictions' closeness to the results in the previous elections may shed light on how much the LLM's predictions are based on past patterns as opposed to new developments. Further, it may be worth exploring whether fine-tuning LLMs on recent, pertinent news and social media debates would improve results, as others have done for BERT with media diets (Chu et al. 2023) or LLaMa with Twitter data (Ahnert et al. 2025).[3] Such content could simply be accessed and analyzed directly, but LLMs may still provide an advantage for aggregating and analyzing such digital trace data.

Our findings are consistent with our hypothesis on contextual biases in an LLM's data-generating process not just vis-à-vis the United States, but also within Europe: GPT-4-Turbo's and Llama's predictions more typically match the voting behavior of Western European countries, which likely can be explained by their larger linguistic and political presence in Europe and presence in the training data. These discrepancies in the digital divide are mirrored in our findings, which suggest that the LLMs are worse at predicting Eastern European voting behavior regardless of prompt language (and information provided in the prompt). The observed ambiguity of prompt language impact on predictive accuracy in our study mirrors existing findings, with some research suggesting prompting in a culture-specific language could mitigate biases to some extent (W. Wang et al. 2024), but other research finding consistent bias across languages (Durmus et al. 2024; Hartmann et al. 2023; Öztürk et al. 2025).

Our use of "context" throughout the study has referred to linguistic and national (political, social, and digital) differences alike. In theory, these are separable mechanisms rather than facets of one underlying construct: unequal linguistic representation in training data, unequal coverage in fine-tuning, and structural mismatches between a country's political system (e.g., multi-party competition) and what an LLM has learned from more heavily represented contexts are each capable of producing the kind of divergences we label "contextual bias", but they call for different remedies. Collapsing them into a single category cannot answer which potential remedy, that is, more training data, more balanced fine-tuning, or a different modeling approach altogether, would actually improve predictive performance for underrepresented populations. While distinguishing them was beyond the scope of this study, we encourage future work on this.

[3] At the time of writing, some studies use such data in LLM-powered public opinion simulations, but they do not fine-tune with it: news diets (Murray and Alexander 2026) or social media data (M. Li and Conrad 2026) is either used to enrich the prompt, or as a direct basis for prediction, with the LLM operating as a classifier (Zhang et al. 2024; Cerina and Duch 2025).

Considering the impact of information contained in the prompt on prediction quality, our findings suggest that demographic information alone is insufficient for accurately estimating complex individual-level attitudes. The fact that providing (general) attitudinal information in the prompt leads to better estimates of voting behavior gives rise to two considerations. It suggests that by adding even more (attitudinal) information about voters (in the European elections case, this might be, e.g., party identification, satisfaction with the national government, salience of and attitude towards issues such as immigration, economic growth, or climate change, and voting behavior in the last election), predictions might further improve (c.f. Gordon et al. 2026; Fröhling et al. 2026; Wolf 2026). However, in our study, such data was not available with the most recent Eurobarometer sample, once again highlighting the tradeoff between recency and detail of human samples on which silicon samples can be based. This leads to the second point: if detailed, up-to-date individual attitudinal information is required for an LLM to make accurate predictions of voting behavior or other items of public opinion, then silicon samples cannot replace surveys, as researchers still need surveys to obtain such information; their potential value rather lies in augmenting existing survey data, for example through imputation or hybrid estimators that combine LLM predictions with smaller human samples (see, e.g., Holtdirk et al. 2025).

Nevertheless, LLM-based predictions have potential for improvement. Future research could benefit from engaging with political scientists' learnings from polling and election forecasting regarding determinants and weights in predictive modeling. In this context, panel survey data might provide additional advantages for research on silicon samples. Comparing our GPT predictions of voting behavior based on pre-election survey data to post-election survey data revealed that the LLM couldn't simply substitute the post-election survey, at least not with a zero-shot, non-fine-tuned approach based on a sample drawn one year ahead of the elections. More accurate predictions might hinge on individual-level pre-election survey data being released in a more timely manner. Experiments with different pre-event waves could reveal where the "survey data cutoff" point is for LLMs to succeed in this task. However, comparing LLM output to survey estimates and treating deviations from those estimates as “bias” presupposes that survey data constitutes the gold standard for measurement quality. Although this is a reasonable starting point, even high-quality survey data is not free from errors, potentially challenging the appropriateness of using it as a benchmark for LLMs. Behavioral or administrative data, such as that which we used as ground truth in this study, could be an alternative, where available. While fine-grained, individual-level observational data (e.g., actual as opposed to reported voting behavior) is hardly available for most social science concepts, our sensitivity analyses revealed that survey data still mirrored the aggregate real-world phenomenon (i.e., election results) better. Future research could evaluate which factors influence the difference of either prediction to such real-world observations. Other conceivable benchmarking alternatives are assessing the belief-structural coherence between surveyed attitudes and simulated behavior rather than agreement with population estimates (e.g., Von der Heyde et al. 2026; Barrie and Cerina 2026), or downstream decision utility for applied use cases – each touching on a different notion of validity.

Regarding the generalizability of our findings to different LLMs, the biases GPT-4-Turbo exhibits are mirrored in the open-weight model Llama. These results suggest a systematic underlying issue in LLM training and fine-tuning that needs to be addressed (McCoy et al. 2024), and highlight the need for better multilingual and multicultural capacities. There are indications that certain models, such as ERNIE (an LLM trained on a balanced mix of English and non-English data) exhibit less cultural bias (W. Wang et al. 2024). Other research suggests that base models are less biased in terms of political orientation, at least on the aggregate level (Rozado 2024), and less sensitive to bias-inducing prompting (Tjuatja et al. 2024) – however, at the cost of less coherence (Rozado 2024). This may suggest that political biases in LLMs are created in the fine-tuning stages, not as a result of biased training data (Rozado 2024). However, politically “neutral” fine-tuning may bring out biases created due to unbalanced training corpora, and even the active alignment against explicit biases might inadvertently exacerbate covert stereotypes (B. Li et al. 2024; Hofmann et al. 2024). Ultimately, “neutrality” is in the eye of the beholder, and alignment processes implicitly mirror the value systems of the people performing the alignment (Kirk et al. 2024). Transparency and diversity in the training and fine-tuning processes can guide the development of fairer and more accurate LLMs for computational social science applications (McCoy et al. 2024; Huckle and Williams 2025). The increased development of LLMs for typically underrepresented languages, such as the TrustLLM (TrustLLM, n.d.), EuroLLM (Martins et al. 2025), or No Language Left Behind (NLLB Team et al. 2024) projects, point in this direction. In addition, since model architecture and training data both influence LLM behavior (McCoy et al. 2024), future research should investigate how different LLMs' outcomes change when provided with different training corpora. However, only a few large companies have the resources to train LLMs, making such experiments largely inaccessible to the scientific community. The importance of an LLM's architecture, and with it, purpose, is evident in our results, where Mistral proved to be entirely unsuitable for the task at hand. This shows that researchers need to seriously consider the intended use cases of off-the-shelf LLMs (McCoy et al. 2024) and potentially customize models for their needs (e.g., Holtdirk et al. 2025). While LLMs may be *general*-purpose tools, that does not mean they are, by default, suitable for highly *specific* tasks such as public opinion prediction.

In conclusion, our findings emphasize the limited applicability of popular LLMs to public opinion prediction. The prediction of attitudes and behaviors relating to events that go beyond LLM training data is what would benefit most from LLMs' efficiency, but such predictions, which necessarily are based on past training and population data, largely fail. Moreover, LLM prediction accuracy is unequally distributed across countries and languages, even when using individual-level prompting information. Finally, improving LLM predictions requires detailed attitudinal information about individuals. Practitioners need to carefully examine the applicability of silicon sampling in the specific target context before drawing any conclusions, so as to not reproduce existing biases. For researchers, our findings point to the need to improve LLMs' training and fine-tuning to mitigate biases and inequalities against specific populations.

## Author Contributions

Leah von der Heyde conceptualized the research design, compiled and summarized the literature, collected, processed, and analyzed the data, and wrote the manuscript.
Anna-Carolina Haensch refined the research design, prepared the automation of data collection, contributed to the literature review, and provided comments and edits to the manuscript.
Alexander Wenz refined the research design, contributed to the literature review, and provided comments and edits to the manuscript.
Bolei Ma selected additional LLMs, collected the LLM-based data for robustness checks, and provided comments and edits to the manuscript.
All authors approved the final version of the manuscript.

## Acknowledgements

The authors would like to thank Pierre Mesure, Monika Sčipova, and Daria Szafran for their support in translating the prompt components to French, Slovak, and Polish. We also thank our student assistants Danylo Syrota and Fedor Miasnikow for supporting the sensitivity analyses.

## Statements and Declarations

### Declaration of Conflicting Interest

The authors declared no conflicts of interest.

### Ethics statement

We use secondary data originally collected by GESIS – Leibniz Institute for the Social Sciences. For more information on informed consent as part of this data collection, see https://doi.org/10.4232/1.14167. No ethical approval was requested for the collection of data generated by large language models.

### Data Availability

This study relies on secondary data originally collected by GESIS – Leibniz Institute for the Social Sciences. This data and corresponding documentation can be downloaded from GESIS via https://doi.org/10.4232/1.14167 upon free-of-charge registration.
Based on this data, synthetic data was generated with LLMs using proprietary APIs. Raw and processed versions of this data are available at https://github.com/leahvdh/United-in-Diversity [repository will be made public upon publication].

## Code Availability

Code allowing for replication of the data collection, processing, and analyses conducted can be downloaded from https://github.com/leahvdh/United-in-Diversity [repository will be made public upon publication].

## Funding

The authors received no financial support for the research, authorship, and publication of this article.

## Appendices

for
**United in Diversity? Contextual Biases in LLM-Based Predictions of the 2024 European Parliament Elections**

**Appendix I:** Prompting (variables, translations, parties, keywords; election results) – see xlsx file

**Appendix II:** Summary statistics per country (prompt variables and number of missing values) – see xlsx file

**Appendix III:** Country selection and additional analyses (GPT, Llama, Mistral) – here below

### 1. Country selection

| Country | Language Internet Coverage | Language Speakers in EU | Language Family | Population Size (millions) | European Region | Political Position within Europe (Shapley-Shubic Index of Bargaining Power) | Share of support for EU member-ship (EB 99.4: QA12.2) |
|---|---|---|---|---|---|---|---|
| **France** | 4.4% | 25% | Romance | 68.2 | West | 13.68 | 55.3% |
| **Germany** | 5.4% | 29% | Germanic | 83.1 | West | 17.23 | 68.9% |
| **Poland** | 1.8% | 9% | Slavic | 36.8 | East | 7.08 | 50.4% |
| **Slovakia** | 0.4% | 2% | Slavic | 5.4 | East | 1.61 | 62.3% |
| **Sweden** | 0.5% | 3% | Germanic | 10.5 | North | 2.39 | 78.5% |

Table A1: *Variation in selected countries' linguistic, geographic, political, and attitudinal contexts.*

## 2. Additional Analyses: Turnout

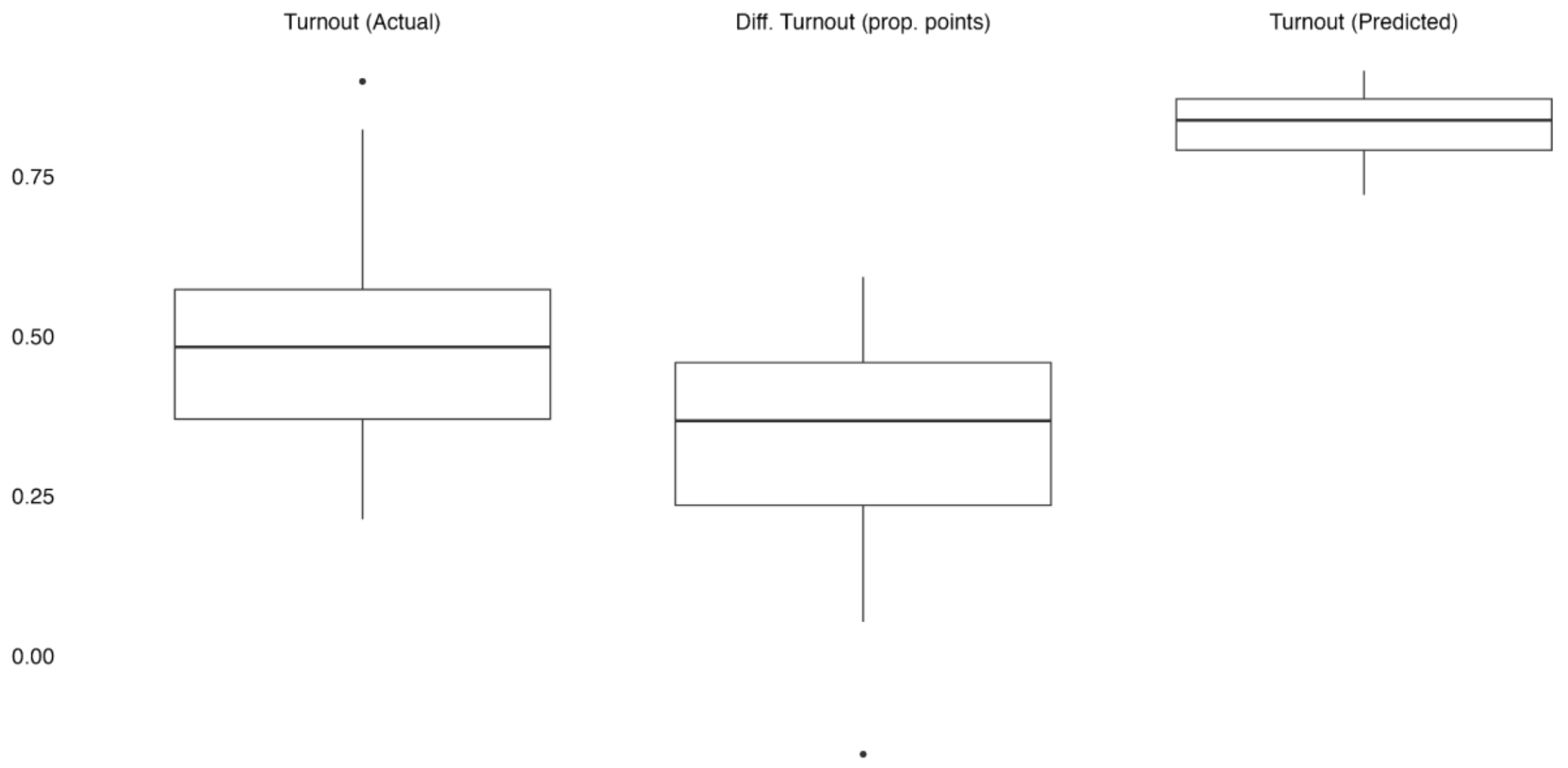

Figure A1: *Distribution of actual and predicted turnout and their differences (based on full English prompt).*

## 3. Additional Analyses: Party Vote Shares

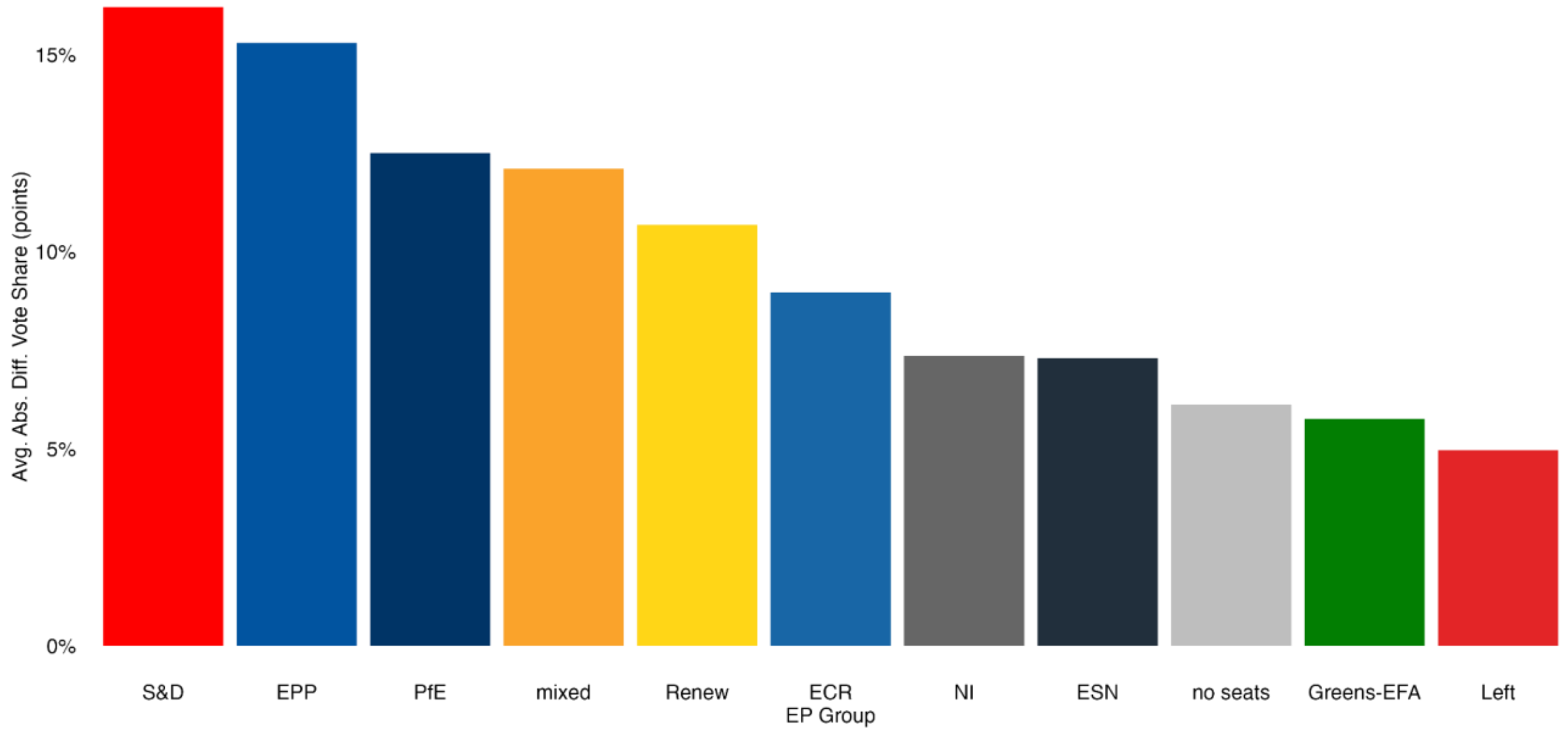

Figure A2: *Average absolute differences in party vote shares by EP group (based on full English prompt).*

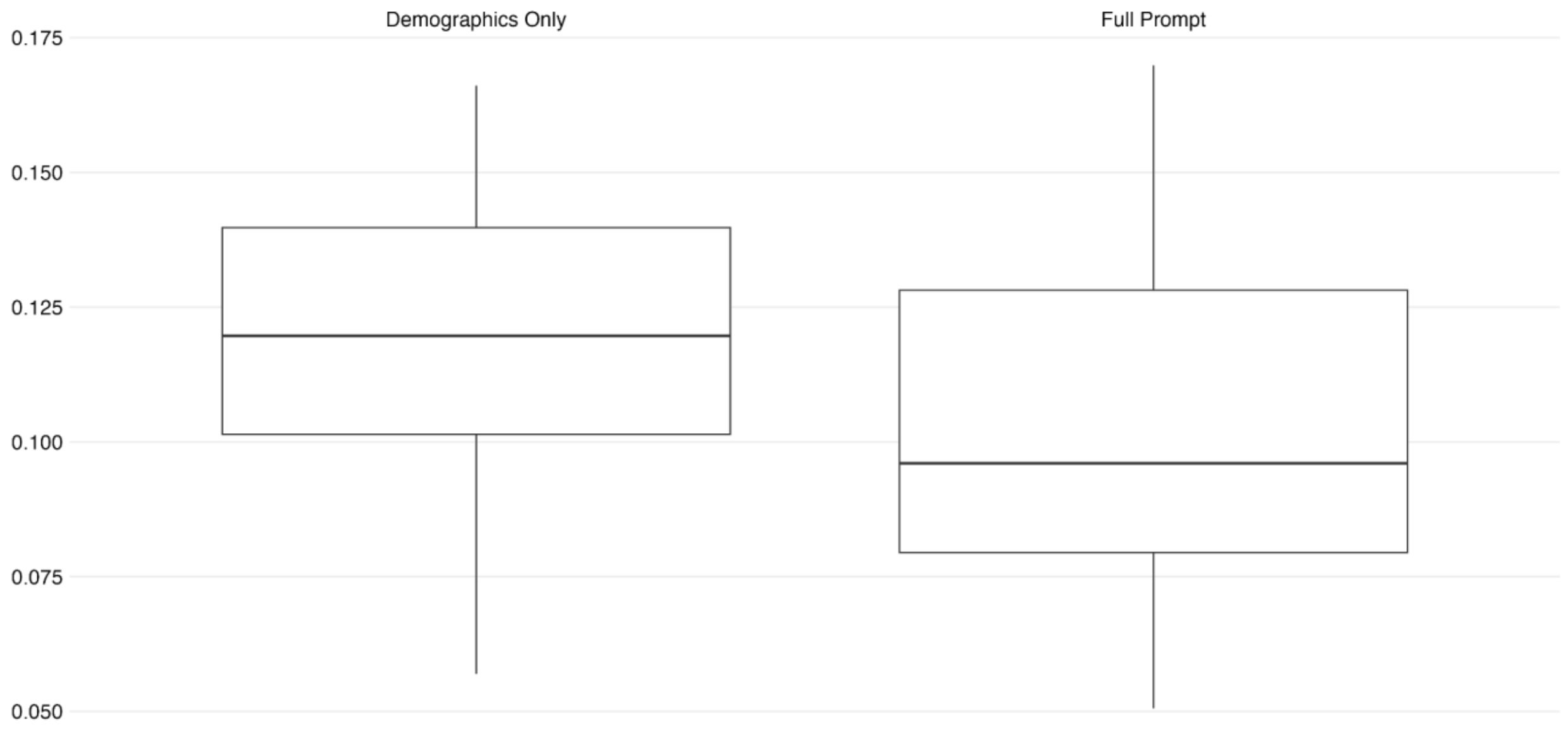

Figure A3: *Distribution of per-country average absolute differences in party vote shares by prompt content, as proportion points (based on English prompt).*

## 4. Analyses for Open-Weight Models

### 4.1. Llama 3.1

Note: Compared to GPT, the cross-country average share of missing data is 114 times higher for predictions of turnout (0.02% vs. 1.8%) and 15 times higher for predictions of party vote shares (0.2% vs. 2.5%). This is the case especially for Poland and Slovakia as well as the other countries whose average includes the completions based on native-language prompts which contained a higher amount of missing values.

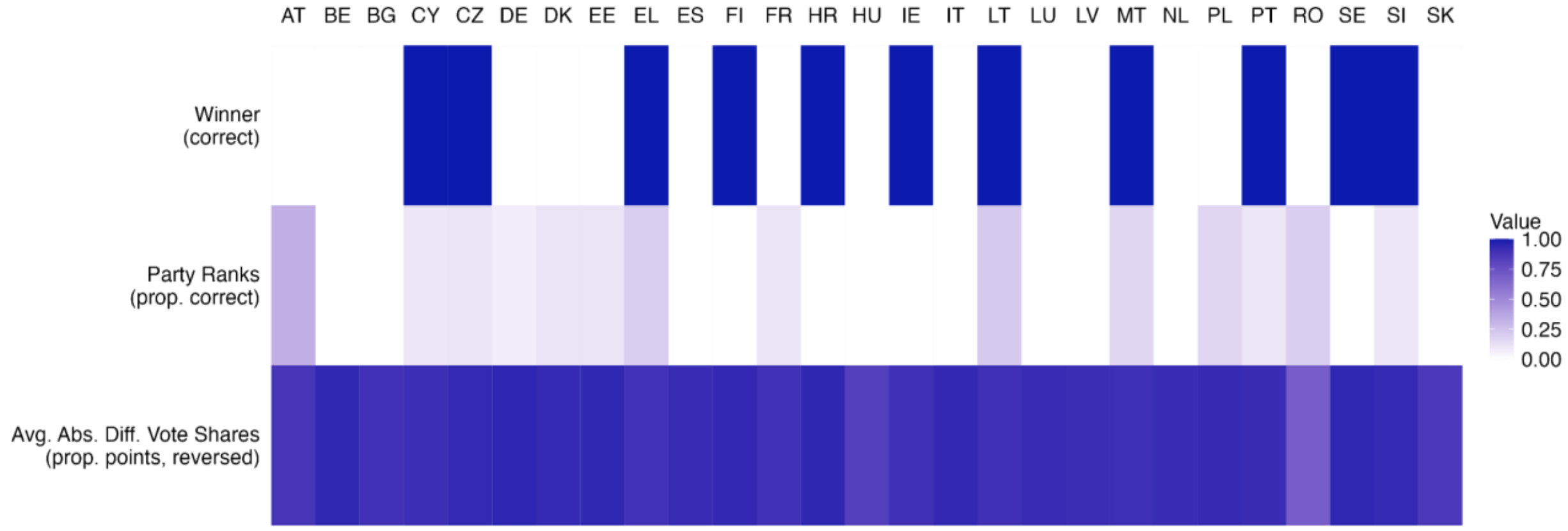

Figure A4: *Predictive performance of Llama 3.1 for the 2024 EU election party results (based on full English prompt).*

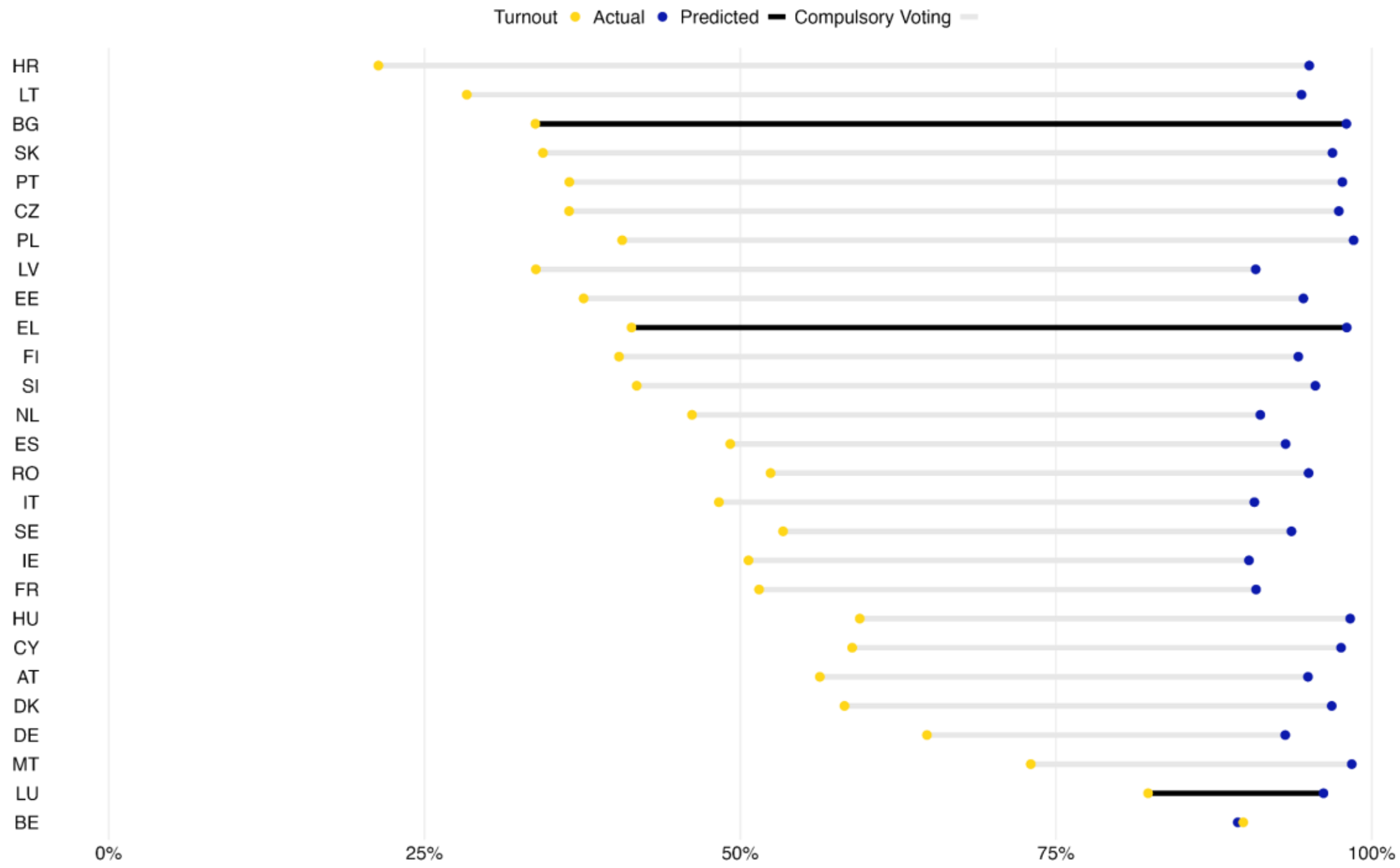

Figure A5: *Difference between actual turnout in the 2024 EU elections and Llama 3.1's predictions (based on full English prompt).*

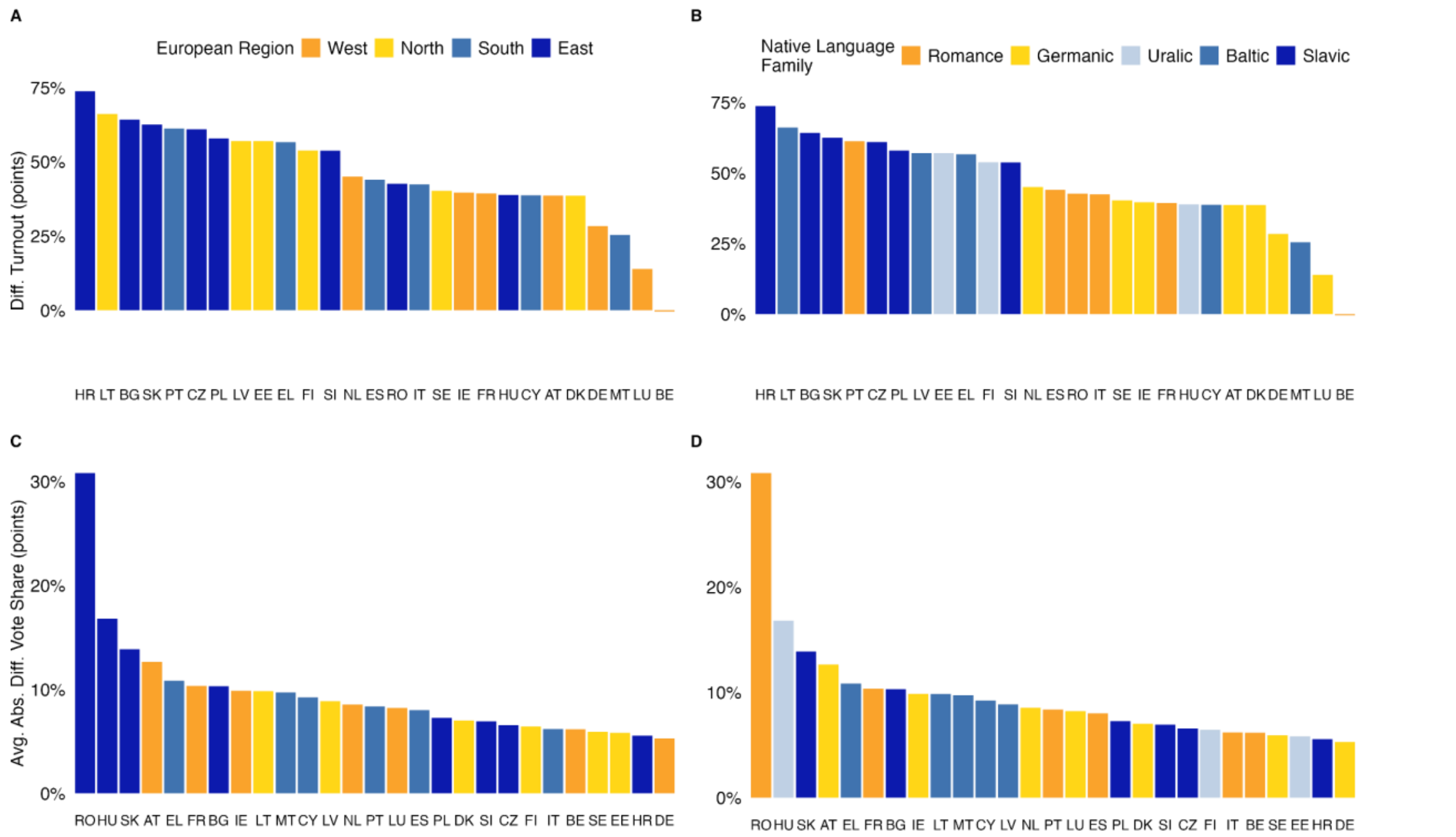

Figure A6: *(Average) difference between actual turnout and party vote shares in the 2024 EU elections and Llama 3.1's predictions (based on full English prompt)* ***by region and language family****.*

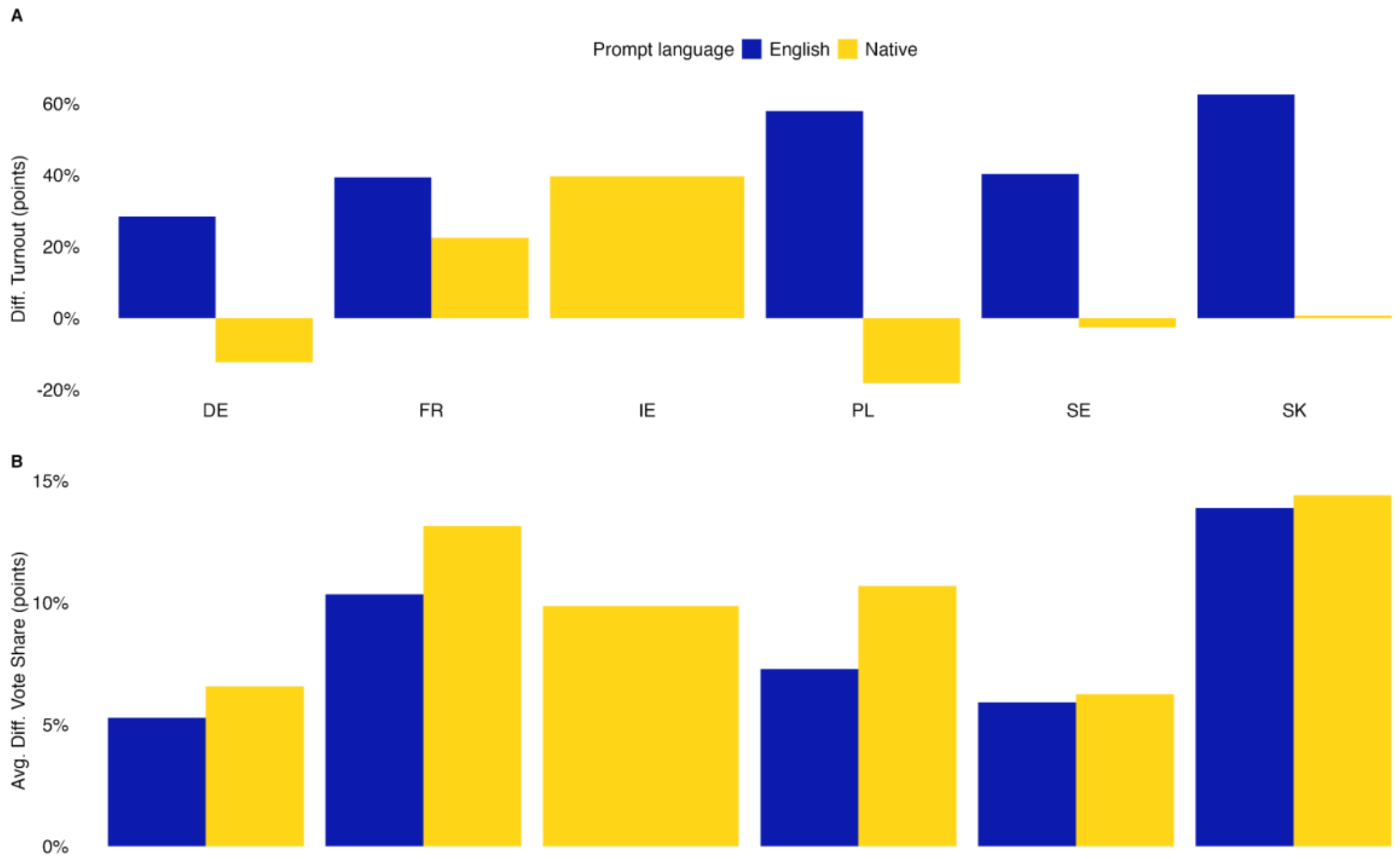

Figure A7: *(Average) difference between actual turnout and party vote shares in the 2024 EU elections and Llama 3.1's predictions (based on full prompt)* ***by prompt language****.*

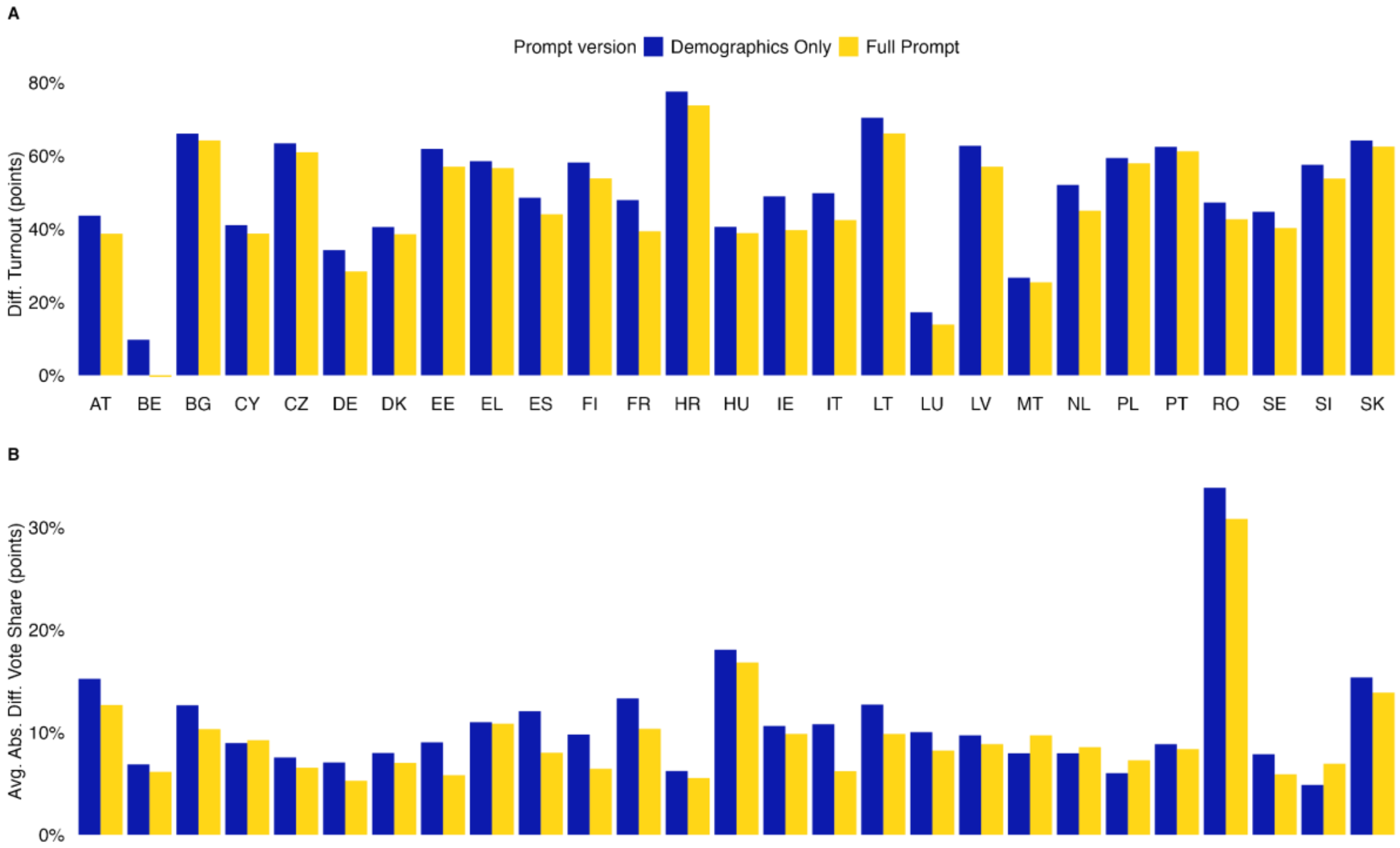

Figure A8: *(Average) difference between actual turnout and party vote shares and predictions using Llama 3.1 (based on* ***English*** *prompt)* ***by prompt content****.*

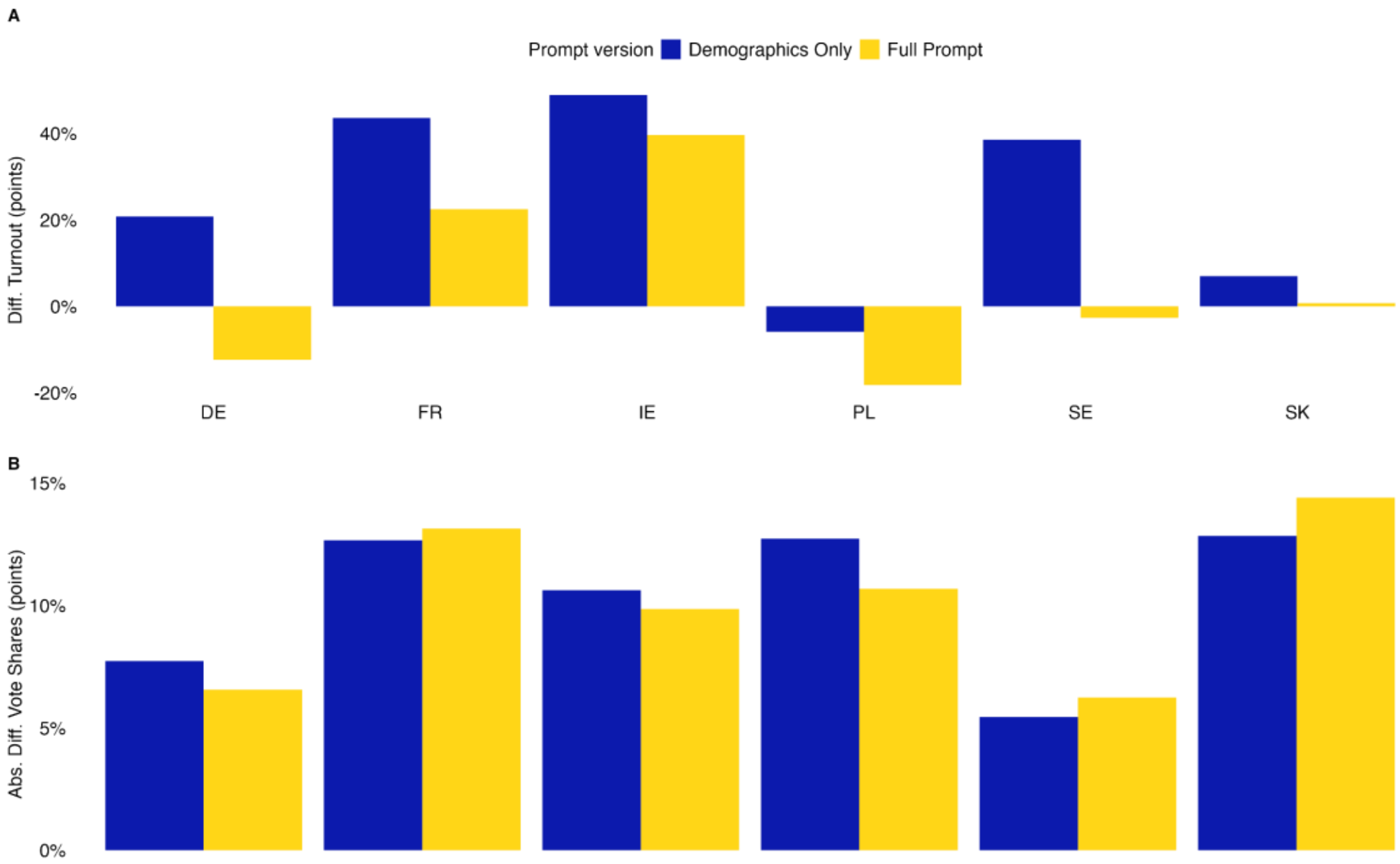

Figure A9: *(Average) difference between actual turnout and party vote shares and predictions using Llama 3.1 (based on* ***native language*** *prompt)* ***by prompt content****.*

### 4.2. Mistral

Manual checks of the automated extractions of vote choices revealed that a disproportionate amount of completions lacked a vote choice. For 11 out of 64 country-language-prompt version-permutations, systematic manual checks of the automated vote choice extraction confirmed the large amount of missing values, upon which it was decided not to pursue further analyses. The sample of manual checks included English and native-language prompting, full-information and demographics-only prompting, Eastern and Western European countries, countries with Slavic, Romance, and Germanic native languages, and countries using Arabic and Cyrillic alphabets, corroborating that this is a general issue with Mistral. Notably, there are fewer missing values in the datasets using German prompting, and more missing values in those using a demographics-only prompt.

| **Country** | **Language** | **Prompt Version** | **Share NAs: Turnout** | **Share NAs: Party Choice (of non-NAs for turnout)** |
|---|---|---|---|---|
| AT | EN | full | 46.2% | 34.4% |
| AT | EN | dem. | 71.8% | 8.8% |
| BE | EN | full | 50.2% | 68.3% |
| BE | EN | dem. | 75.2% | 66.4% |
| BG | EN | full | 41.8% | 46.2% |
| BG | EN | dem. | 72.9% | 18.2% |
| CY | EN | full | 55.0% | 77.4% |
| CY | EN | dem. | 69.5% | 46% |
| CZ | EN | dem. | 77.8% | 45.6% |
| DE | DE | full | 0.9% | 14.3% |
| DE | DE | dem. | 3.0% | 16.8% |

Table A2: *Proportions of missing values in selected datasets using Mistral.*

## 5. Comparison to Post-Election Survey Data

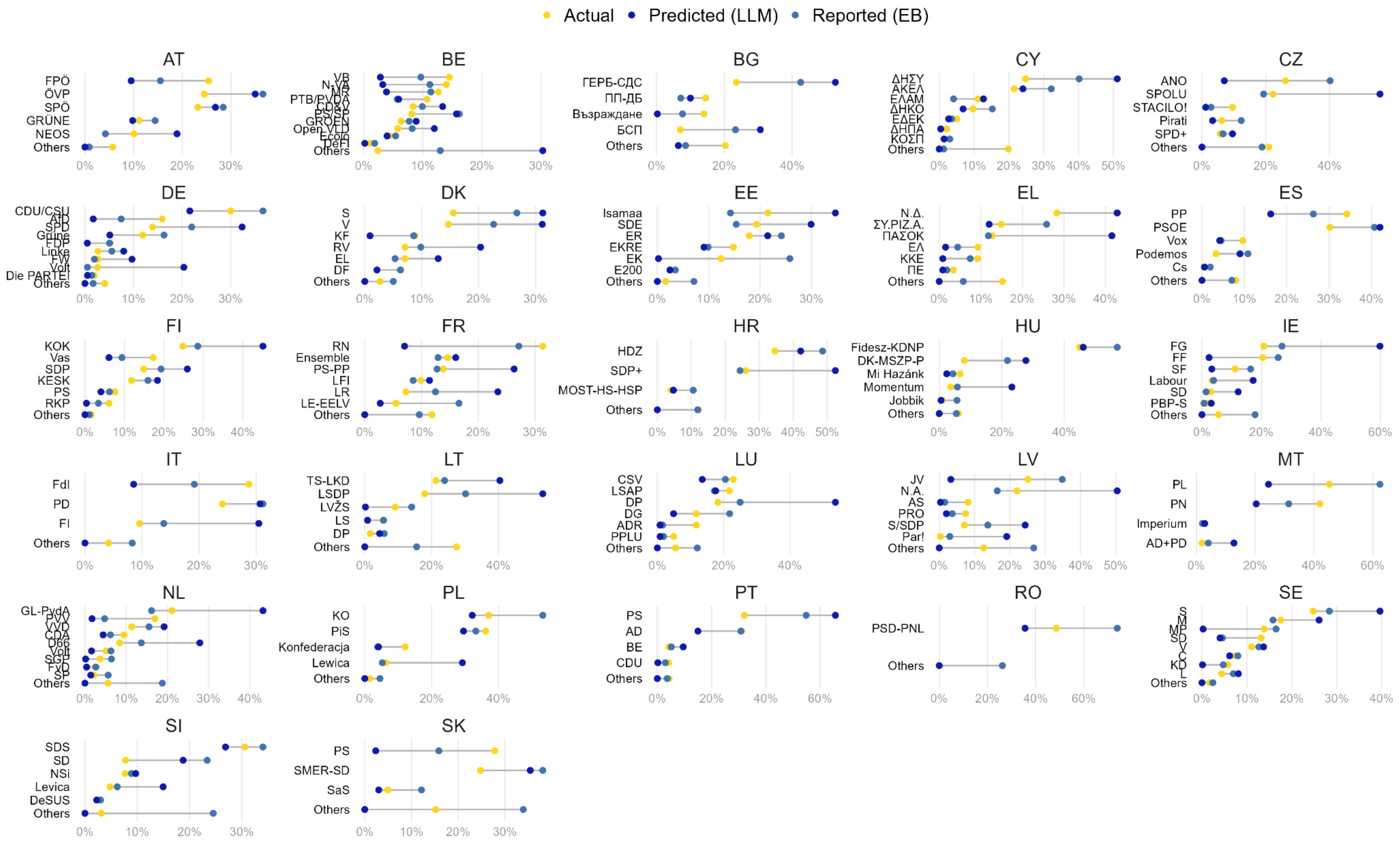

Figure A11: *Differences between actual and predicted party vote shares (GPT-4-Turbo based on full English prompt and post-election Eurobarometer) by country and party.*

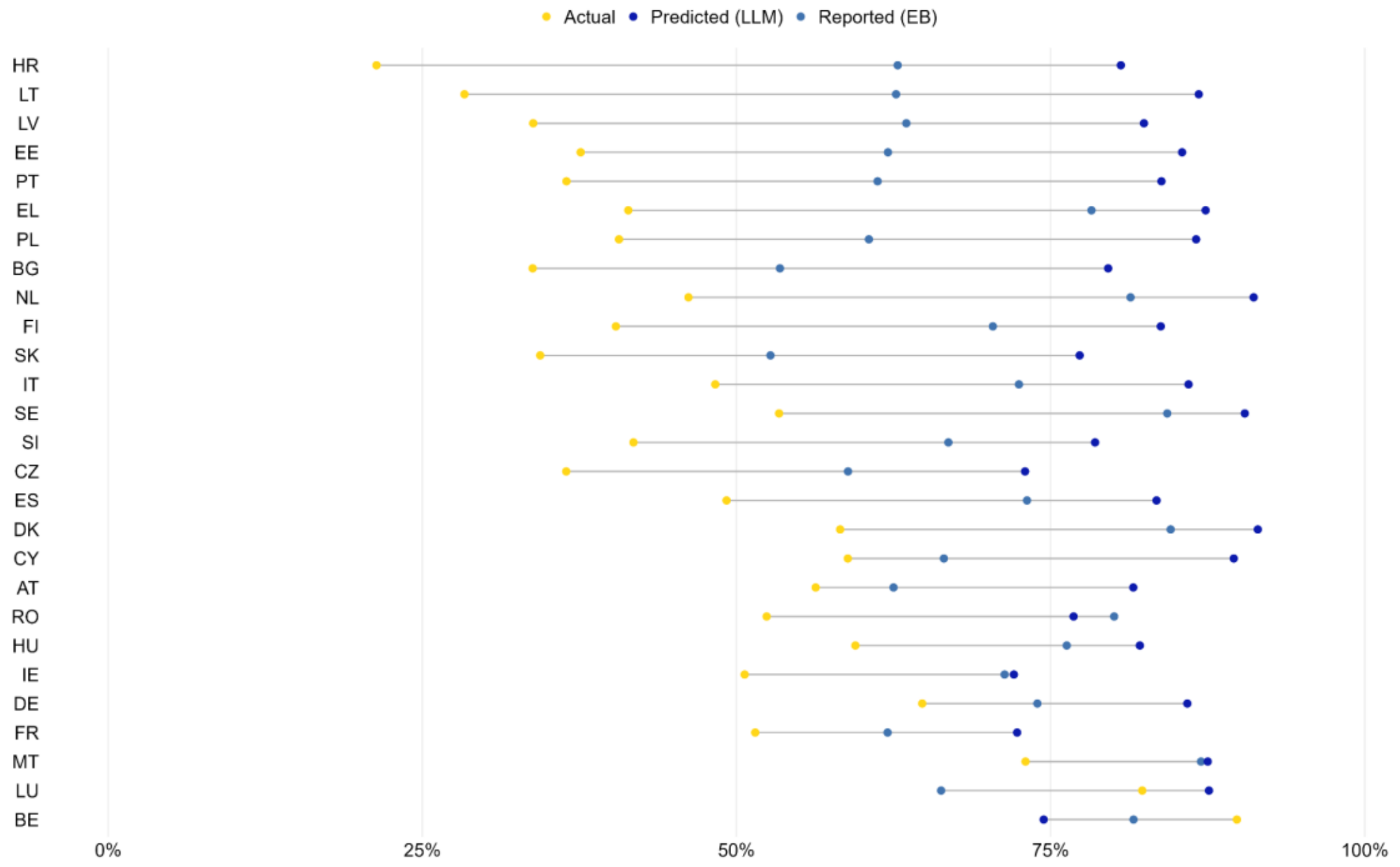

Figure A11: *Difference between actual turnout in the 2024 EU elections, the post-election survey estimate, and GPT-4-Turbo's predictions (based on full English prompt) by country.*

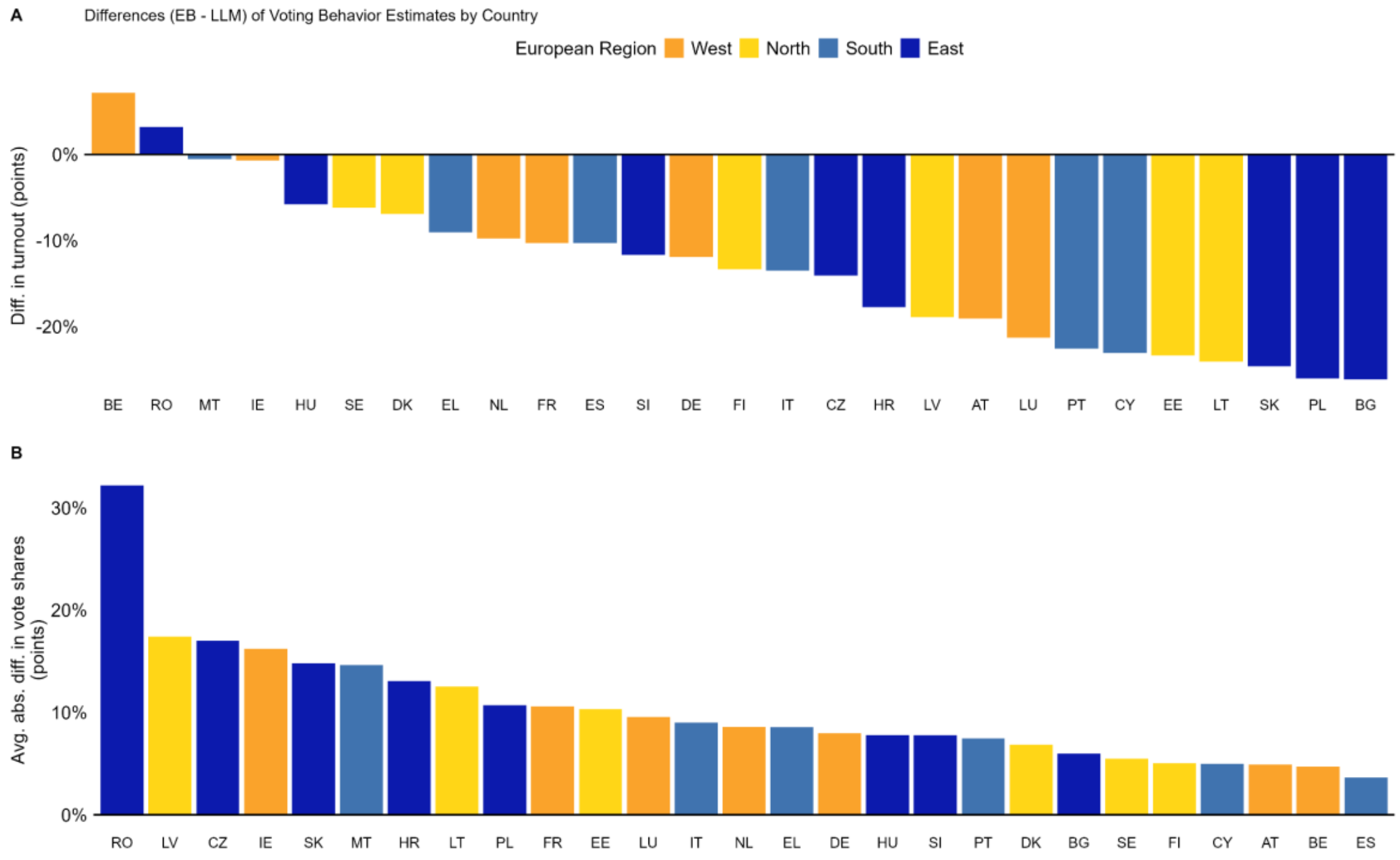

*Figure A12: (Average) difference between turnout and party vote shares predicted by GPT-4-Turbo's (based on full English prompt) and reported in the post-election Eurobarometer* ***by region.***